\documentclass[a4paper, preprint, 12pt, aps, prb, amsmath, amssymb]{revtex4-1}
\usepackage[utf8]{inputenc}
\usepackage{graphicx}
\usepackage{bm}

\newcommand{\vrad}{\ensuremath{v_\mathrm{rad}}}
\newcommand{\neng}{\ensuremath{\overline{n}_\text{e} / n_\text{G}}}
\newcommand{\Isat}{\ensuremath{I}}

\newcommand{\Te}{\ensuremath{T_\mathrm{e}}}
\newcommand{\mi}{\ensuremath{m_\mathrm{i}}}

\newcommand{\veff}{\ensuremath{v_\mathrm{eff}}}

\newcommand{\cs}{\ensuremath{C_{\mathrm{s}}}}
\newcommand{\Ip}{\ensuremath{I_\mathrm{p}}}
\newcommand{\vthe}{\ensuremath{v_{\mathrm{th}, \mathrm{e}}}}
\newcommand{\tauw}{\ensuremath{\tau_{\mathrm{w}}}}
\newcommand{\taud}{\ensuremath{\tau_{\mathrm{d}}}}
\newcommand{\taur}{\ensuremath{\tau_{\mathrm{r}}}}
\newcommand{\tauf}{\ensuremath{\tau_{\mathrm{f}}}}

\newcommand{\smean}[1]{\ensuremath{\overline{ #1 }}}
\newcommand{\mean}[1]{\ensuremath{\langle #1 \rangle}}
\newcommand{\rms}[1]{\ensuremath{{ #1 }_\text{rms}}}
\newcommand{\bigexp}[1]{\ensuremath{\exp \left( #1 \right)}}

\newcommand{\musec}{\ensuremath{\mu \mathrm{s}}}
\newcommand{\Figref}[1]{Fig.~\ref{fig:#1}}
\newcommand{\Figsref}[1]{Figs.~\ref{fig:#1}}

\newcommand{\Tabref}[1]{Tab.~\ref{tab:#1}}
\newcommand{\tabref}[1]{tab.~\ref{tab:#1}}
\newcommand{\Eqsref}[1]{Eqs.~\eqref{eq:#1}}
\newcommand{\Eqnref}[1]{Eqn.~\eqref{eq:#1}}

\newcommand{\Secref}[1]{Section \ref{sec:#1}}

\newcommand{\Ref}[1]{[\onlinecite{#1}]}

\begin{document}
\title{Fluctuation statistics in the scrape-off layer of Alcator C-Mod}
\author{R. Kube}
\email[E-mail:]{ralph.kube@uit.no}
\author{A. Theodorsen}
\author{O.E. Garcia}
\affiliation{Department of Physics and Technology,
UiT - The Arctic University of Norway,
N-9037 Tromsø, Norway}
\author{B. LaBombard}
\author{J.L. Terry}
\affiliation{MIT Plasma Science and Fusion Center, Cambridge, MA, 02139, USA}

\date{\today}
\begin{abstract}
We study long time series of the ion saturation current and floating potential,
sampled by Langmuir probes dwelled in the outboard mid-plane scrape off layer and 
embedded in the lower divertor baffle of Alcator C-Mod. A series of ohmically heated 
L-mode plasma discharges is investigated with line-averaged plasma density ranging from
$\neng = 0.15$ to $0.42$, where $n_\mathrm{G}$ is the Greenwald density.
All ion saturation current time series that are sampled in the far scrape-off layer are 
characterized by large-amplitude burst events. Coefficients of skewness and excess 
kurtosis of the time series obey a quadratic relationship and their histograms 
coincide partially upon proper normalization. Histograms of the ion saturation current 
time series are found to agree well with a prediction of a stochastic model for the 
particle density fluctuations in scrape-off layer plasmas.

The distribution of the waiting times between successive large-amplitude burst events and of 
the burst amplitudes are approximately described by exponential distributions. The average 
waiting time and burst amplitude are found to vary weakly with the line-averaged plasma density.
Conditional averaging reveals that the radial blob velocity, estimated from floating potential
measurements, increases with the normalized
burst amplitude in the outboard mid-plane scrape-off layer. For low density discharges,
the conditionally averaged waveform of the floating potential associated with large
amplitude bursts at the divertor probes has a dipolar shape. In detached divertor conditions
the average waveform is random, indicating electrical disconnection of blobs from the
sheaths at the divertor targets.
\end{abstract}

\maketitle

\section{Introduction}
\label{sec:introduction}
The far scrape-off layer of magnetically confined plasmas is dominated by intermittent 
fluctuations of the particle density and concomitant large transport events.
A large body of research links these phenomena to the radial propagation
of plasma filaments which are elongated along the magnetic field and localized in the
radial-poloidal plane. 
\Ref{wootton-1990, krash-2001, boedo-2001, zweben-2002, boedo-2003, rudakov-2005, zweben-2007, garcia-2009, dippolito-2011}.
They carry excess particle density and heat relative to the background plasma and are hence 
called \emph{blobs}
Blobs are believed to mediate the parallel and perpendicular transport channels of particle and 
heat fluxes in the scrape-off layer. Furthermore may blob propagation be responsible for a significant 
heat load on plasma facing components of the vacuum vessel. As the empirical discharge density limit
\Ref{greenwald-1988} is approached, the relative magnitude of these transport channels changes such 
as to favor perpendicular transport \Ref{labombard-2001, garcia-2007-jnm}. 
To understand the impact of blobby transport on plasma confinement, their mode of propagation as well 
as the statistics of fluctuation induced transport have to be studied.

%%%%%%%%%%%%%%%%%%%%%%%%%%%%%%%%%%%%%%%%%%%%%%%%%%%%%%%%%%%%%%%%%%%%%%%%%%%%%%%%%%%%%%%%%
% Effects of blobs:
% 1.) Generation: Mechanisms, rate, dependency on plasma parameters
% 2.) Once generated, are they stable? 
% 3.) How far to stable blobs travel. And how fast? ( -> Copmetition between parallel and perp transport)
%To assess the importance of plasma blobs to plasma confinement, questions regarding
%the generation mechanism of plasma blobs, their stability, and their mode of propagation
%have to be addressed.
%It has been suggested that large amplitude drift-wave turbulence events that have become
%unstable may lead to blob events \Ref{krash-2007}, which has been observed in the TJ-K
%stellarator \Ref{furchert-2013}.

The basic mechanism underlying plasma blob propagation is the interchange mechanism. 
\Ref{krash-2001, bian-2003, garcia-2006-ps, garcia-2006, theiler-2009, kube-2011}. 
Magnetic gradient and curvature drifts in an inhomogeneous magnetic field 
give rise to an electric current which polarizes filament structures of elevated pressure 
perpendicular to the magnetic field and its direction of variation. At the outboard 
mid-plane location of a toroidally magnetized plasma, a filament of elevated pressure 
is polarized in such a way that it propagates radially outwards towards the main chamber
wall \Ref{krash-2001}.

The path along which electric currents within the filament are closed are crucial for 
its radial velocity. Assuming that the electric current in the plasma filament can flow freely
along magnetic field lines within the plasma filament, the electric current loop may be 
closed through these sheaths. Two-dimensional fluid simulations of isolated plasma blob 
propagation reveal that the radial blob velocity decreases with increasing magnitude of 
the parallel electric currents, parameterized by a sheath dissipation parameter 
\Ref{krash-2001, garcia-2006, kube-2011}. Sheath connection is expected to be limited by 
ballooning of the plasma filaments and large plasma resistivity which prevents parallel 
electric currents through the sheaths \Ref{myra-2006, russell-2007, easy-2014, easy-2016}.
Fluid modeling of plasma blobs furthermore reveals a dependence of its radial velocity, 
$\vrad$, on the relative blob amplitude, where blobs with larger amplitude feature a larger
radial center of mass velocity \Ref{kube-2013, wiesenberger-2014, angus-2014}.

Studies of plasma blob propagation in Alcator C-Mod show a good agreement between their
radial velocity and the sheath-connected velocity scaling law when the scrape-off layer
is sheath-limited \Ref{kube-2013}.
Work at Alcator C-Mod furthermore reveals correlation coefficients of up to 75\% between time 
series of particle density proxies, sampled at different poloidal positions along a single 
magnetic field line \Ref{grulke-2006, grulke-2014}. This supports the idea that blobs in Alcator C-Mod 
may extend from the outboard mid-plane to the divertor sheaths and are sheath connected in suitable low-density
plasmas.

%%%%%%%%%%%%%%%%%%%%%%%%%%%%%%%%%%%%%%%%%%%%%%%%%%%%%%%%%%%%%%%%%%%%%%%%%%%%%%%%%%%%%%%%%
%
% Statistical features
%
The turbulent flows in the far scrape-off layer of magnetically confined plasmas have 
been demonstrated to have many universal properties 
\Ref{antar-2001, antar-2003, vanmilligen-2005, graves-2005, horacek-2005}.
For one, the conditionally averaged waveform of large amplitude events in particle density 
time series presents a fast rise and a slow fall 
\Ref{boedo-2001, rudakov-2002, boedo-2003, kirnev-2004, rudakov-2005, xu-2005, 
     garcia-2006-tcv, garcia-2007-tcv, garcia-2007-coll, boedo-2014, garcia-2015}.
The conditionally averaged waveform as well as the histogram of ion saturation current 
time series were found to collapse upon normalization for a range of line-averaged 
plasma densities and plasma currents in the \emph{Tokamak \`a configuration variable} 
(TCV) tokamak \Ref{garcia-2007-tcv, garcia-2007-coll, garcia-2009}. Exponentially 
distributed burst amplitudes and waiting times have been observed in the scrape-off layer 
of Alcator C-Mod and TCV \Ref{garcia-2013, garcia-2015}. Correlation analysis further 
reveals the presence of a dipolar electric potential structure centered around local 
maxima of the particle density \Ref{grulke-2006, devynck-2006, carter-2006, theodorsen-2016}.

Time series with frequent large amplitude bursts feature histograms with elevated tails
as well as positive coefficients of sample skewness and excess kurtosis
\Ref{antar-2001prl, antar-2001, sattin-2004, graves-2005}. 
The universal character of the fluctuations manifests itself in the fact that histograms 
of the particle density coincide upon normalization when obtained at a single position 
in the far scrape-off layer for various plasma parameters 
\Ref{antar-2001, sattin-2004, graves-2005, vanmilligen-2005, garcia-2007-tcv}.
%Optical measurements of the particle density fluctuations in the scrape-off layer of Alcator 
%C-Mod show good agreement with a Gamma distribution over almost four decades in normalized 
%probability density \Ref{garcia-2013}. 
Particle density fluctuations in the scrape-off layer sampled at the TCV device were also 
found to be well described by a Gamma and a log-normal distribution over a large range of 
discharge parameters \Ref{graves-2005}. 

Another salient feature of the density time series is a quadratic relation between sample 
skewness, $S$, and excess kurtosis, $F$, of the form $F = a  + b S^2$, where $a$ and $b$ 
are real coefficients \Ref{graves-2005, labit-2007, sattin-2009}. This relation is 
intrinsic to some probability distribution functions that have been proposed to describe 
histograms of the particle density time series. Data sampled in the TORPEX device 
over a large range of discharge conditions and spatial locations was shown to be well 
described by the generalized beta distribution \Ref{labit-2007}. Recent work models 
particle density time series as a stochastic process which is based on the superposition 
of individual pulses \Ref{garcia-2012}. 
Under the assumption that the individual pulses decay exponentially, have exponentially
distributed amplitudes and waiting times between pulses, this model predicts the 
particle density time series to be Gamma distributed. It was shown that this 
model describes the intensity fluctuations at the outboard mid-plane scrape-off layer of
Alcator C-Mod, as measured by gas-puff imaging, over several decades in normalized
probability \Ref{garcia-2013}.
So far however, no consensus on one particular analytic model exists in the fusion community.
%
%%%%%%%%%%%%%%%%%%%%%%%%%%%%%%%%%%%%%%%%%%%%%%%%%%%%%%%%%%%%%%%%%%%%%%%%%%%%%%%%%%%%%%%%%

In this paper, we present an analysis of long time series of the ion saturation current and 
floating potential obtained by Langmuir probes in the boundary region of the Alcator C-Mod tokamak.
Utilizing a probe dwelled in the outboard mid-plane scrape-off layer as well as probes 
embedded in the divertor baffle allows us to study the universality of the fluctuations sampled
at these two positions as well as the dependence of the statistics on the line-averaged plasma density.

The structure of this article is as follows. \Secref{fluct_stat} introduces 
a stochastic model for density fluctuations in the scrape-off layer as well as the conditional
averaging method to be used. The experimental setup is described in \Secref{setup}.
\Secref{sol_results} presents the time series analysis data sampled by the probe in the outboard
mid-plane scrape-off layer and \Secref{div_fluct} presents the corresponding analysis 
of the time series obtained from the divertor probes. A discussion of the universal properties
of the time series sampled in both positions and their relation to blob theory are given in
\Secref{discussion}. We conclude in \Secref{conclusion} with suggestions for further work.
%%%%%%%% Correlation of radial velocity and burst amplitude %%%%%%%%%%%%%%%

\section{Fluctuation statistics}
\label{sec:fluct_stat}
%%%%%%%%%%%% Time series: Fit PDF on various models:
Recent work models the particle density fluctuations at a single point in scrape-off layer
plasmas as the superposition of random pulse events 
\Ref{garcia-2012}:
\begin{align}
    \Phi(t) & = \sum_{k} A_k \phi(t - t_k). \label{eq:shotnoise_superposition}
\end{align}
Given that the arrival of pulses in the time series is governed by a Poisson process, 
this model predicts a quadratic relation between coefficients of skewness and excess 
kurtosis. Moreover, by assuming a Poisson distribution it follows that the waiting time 
between pulses are exponentially distributed. Further assuming an exponential pulse shape, 
$\phi(t) = \Theta(t) \exp \left( -t / \taud \right)$, where $\Theta$ is the Heaviside 
step function and $\taud$ the duration time of a pulse, and exponentially distributed pulse 
amplitudes $A_k$, the model implies that the particle density is Gamma distributed 
\Ref{garcia-2012}. The ratio of pulse duration time and waiting times, 
$\gamma = \taud / \tauw$, is the shape parameter of the Gamma distribution.

To include random fluctuations of the background particle density we add normal 
distributed noise to the signal \Eqnref{shotnoise_superposition}, 
\begin{align}
    \Phi'(t) & = \Phi(t) + N(t). \label{eq:noisy_sn}
\end{align}
Here the normal distributed noise $N$ has vanishing mean and variance $\sigma^2$. The resulting 
probability density function of the random variable is then given by the convolution 
of a $\Gamma$-distribution and a normal distribution and can be written using two 
parameters: $\gamma$ and $\epsilon$. While $\gamma$ relates to the ratio 
of pulse decay and waiting time as before, in other words the density of pulse arrivals, 
$\epsilon$ relates the variance of $N$ to the root mean square value of $\Phi$ via 
$\sigma^2 = \epsilon \rms{\Phi}^2$. A large value of $\epsilon$ denotes the case where 
the root mean square value of the process $\Phi$ is smaller than the scale 
parameter of the white noise $N$ and a small value of $\epsilon$ denotes the converse relation.

It is commonly observed in particle density fluctuation time series in scrape-off layer 
plasmas that pulses overlap as to form large amplitude burst events. To determine the 
average structure of the bursts we employ conditional averaging \Ref{pecseli-1989}. 
Starting from the largest burst event in the time series at hand, we identify a set of 
disjunct sub records, placed symmetrically around the peak of burst events which exceed a 
given amplitude threshold until no more burst events exceeding this threshold are left 
uncovered. The threshold is often chosen to be $2.5$ times the root mean square value of 
the time series at hand. This average can be written as
\begin{align}
    C(\tau) & = \langle \Phi(\tau) | \Phi(\tau=0) > 2.5\, \rms{\Phi}\rangle \label{eq:cond_avg},
\end{align}
where $\tau$ is the time offset relative to the peak of the burst. 
The variability of the burst events is characterized by the conditional variance
\Ref{oynes-1995}:
\begin{align}
    %\mathrm{cvar}(\Phi) & = \frac{\langle \left( \Phi - \langle \Phi | C \rangle \right)^2 | C \rangle} {\langle \left( \Phi | C \right)^2 \rangle }. \label{eq:cond_var}
    1 - \mathrm{CV}(\tau) & = 1 - \frac{\langle \left( \Phi - C \right)^2 | \Phi(0) > 2.5\, \rms{\Phi} \rangle}{C^2}.
\end{align}
This quantity is bounded, $0 < 1 - \mathrm{CV}(\tau) < 1$, where the values $0$ and $1$
indicate respectively no and perfect reproducibility of the conditionally averaged 
waveform.

To study the intermittency of ion saturation current time series, they are rescaled
according to
\begin{align}
    \widetilde{\Isat} = \frac{\Isat - \smean{\Isat}_\mathrm{mv}}{I_\mathrm{rms, mv}}. \label{eq:is_normalization}
\end{align}
The subscripts $\mathrm{mv}$ and $\mathrm{rms, mv}$ denote the moving average and moving root 
mean square value respectively. Both are computed within a window
of $16384$ elements when applied to data from the horizontal scanning probe.
This window corresponds to roughly $3 \mathrm{ms}$ and exceeds typical autocorrelation
times of approximately $15\, \mu \mathrm{s}$ by a factor of $200$ \Ref{labombard-2001}. The
same window length is used for the time series obtained by the divertor
probes. In the latter case, this corresponds to approximately $20\, \mathrm{ms}$. Since the
amplitude of the density fluctuation in the scrape-off layer is sensitive to the distance
to the last closed flux surface we compute the statistics within a moving window as to alleviate
the fluctuations of the last closed flux surface indicated in \Figref{probes_rho}. The use of such 
averaging has little influence on the conditional averaging threshold \Eqnref{cond_avg}.
Time series of the floating potential are rescaled by removing a linear trend 
from the time series and subsequently normalizing the time series to the electron
temperature and as to have vanishing mean:
\begin{align}
    \widetilde{V} = \frac{e \left(V - \smean{V} \right)}{T_\mathrm{e}}. \label{eq:vf_normalization}
\end{align}
We do not use a moving average for the floating potential since the amplitude of the signal
varies little with distance to the last closed flux surface.

%%%%%%%%%%%%%%%%%%%%%%%%%%%%%%%%%%%%%%%%%%%%%%%%%%%%%%%%%%%%%%%%%%%%%%%%%%%%%%%%%%%%%%%%%%%%%%%%%%%
%%%%%%%%%%%%%%%%%%%%%%%%%%%%%%%%%%%%%%%%%%%%%%%%%%%%%%%%%%%%%%%%%%%%%%%%%%%%%%%%%%%%%%%%%%%%%%%%%%%
%%%%%%%%%%%%%%%%%%%%%%%%%%%%%%%%%%%%%%%%%%%%%%%%%%%%%%%%%%%%%%%%%%%%%%%%%%%%%%%%%%%%%%%%%%%%%%%%%%%
%
\section{Experimental setup}
\label{sec:setup}
\begin{figure}[htb]
    \begin{minipage}{\textwidth}
        \includegraphics[width=8.5cm]{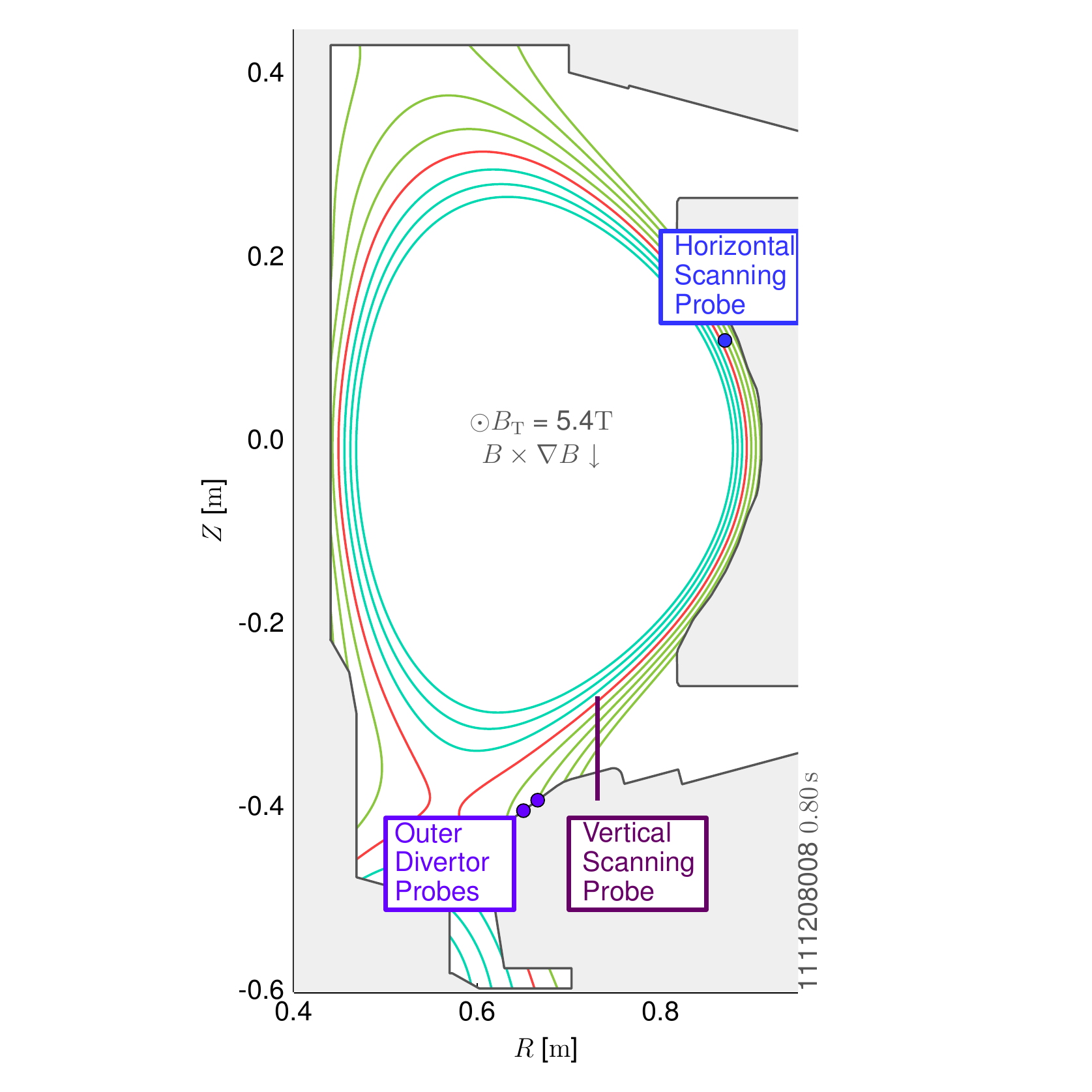}
        \caption{Cross-section of Alcator C-Mod showing the diagnostics from which we report measurements: The horizontal scanning
            probes as well as two probes of the outer divertor probe array. Overlaid are magnetic field lines from discharge 2, 
            as reconstructed by EFIT.}
        \label{fig:acm_cross}
    \end{minipage}
\end{figure}

\begin{figure}[htb]
    \begin{minipage}{\textwidth}
        \includegraphics[width=8.5cm]{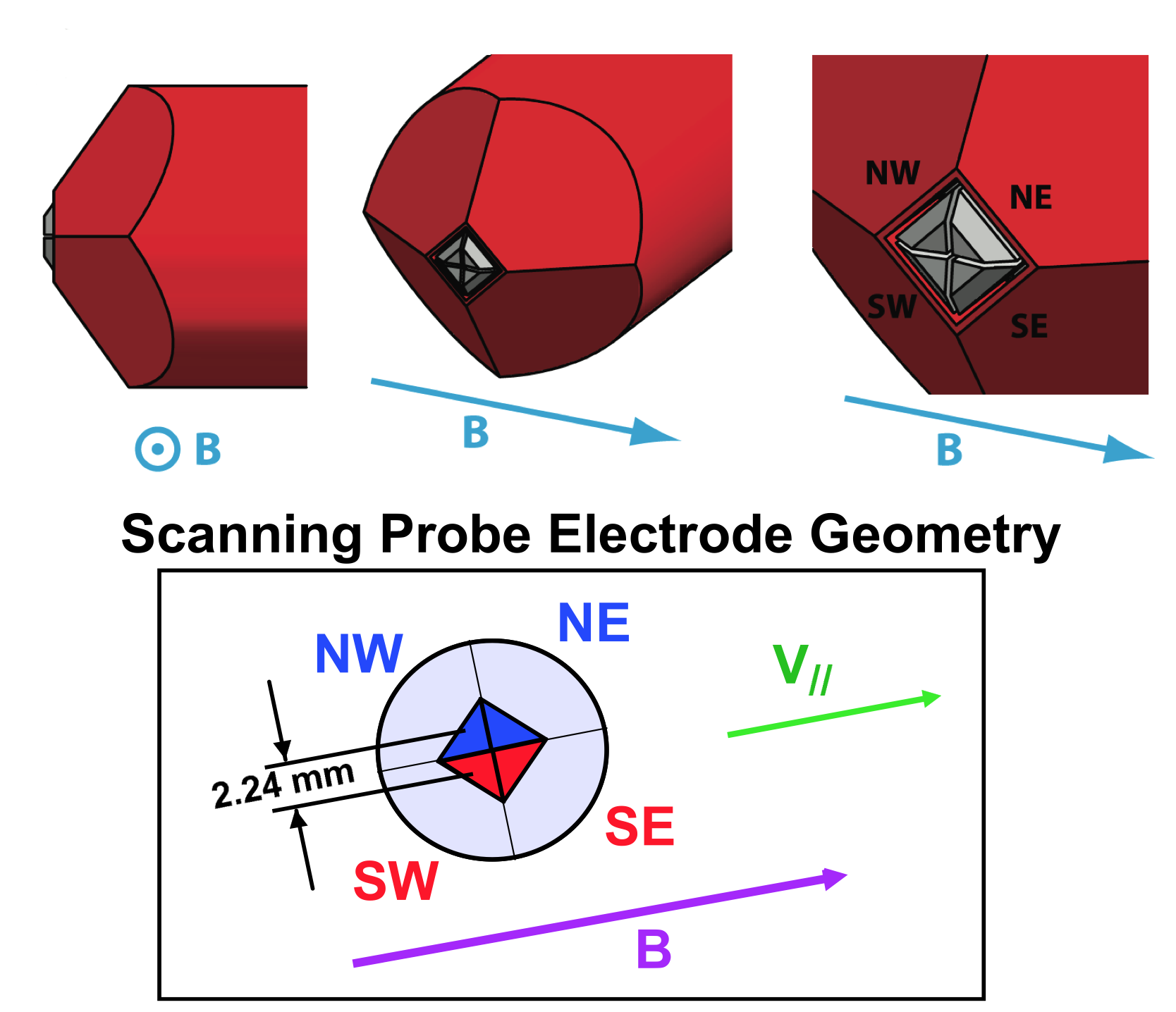}
        \caption{Mach probe head installed on the horizontal and vertical scanning probes.}
        \label{fig:mach_head}
    \end{minipage}
\end{figure}

%%% Overview of the facility
Alcator C-Mod is a compact tokamak with a major radius of $R=0.68\,\mathrm{m}$ and
a minor radius of $a=0.22\,\mathrm{m}$, and allows for a magnetic field of up to
$8 \mathrm{T}$ on-axis.
Figure \ref{fig:acm_cross} shows a cross-section of Alcator C-Mod together
with the diagnostics from which we report measurements: the horizontal and vertical
scanning probes and the Langmuir probe array embedded in the lower outer divertor 
baffle. The magnetic equilibrium from discharge 2 of \Tabref{shotparams}, as reconstructed 
by EFIT \Ref{lao-1985}, is overlaid.
%%%%%%% Describe the probe head we are using.
The Mach probe head installed on both scanning probes, depicted in \Figref{mach_head},
is designed to routinely 
withstand heat fluxes of up to $100\, \mathrm{MW} / \mathrm{m}^2$ \Ref{smick-2009, smick-2013}.
All electrodes are connected to sampling electronics that sample current and voltage
with $5\, \mathrm{MHz}$ and $14$ bit resolution.
% Where are the probes
The horizontal scanning probe is installed $10\, \mathrm{cm}$ above the outboard
mid-plane and can be reciprocated horizontally $11 \, \mathrm{cm}$ into the plasma.
For the present experiments this probe was dwelled at a fixed position in the scrape-off
layer for the entire duration of the plasma discharge. 
% Map shit to one thing, yo
As a common radial coordinate we employ the magnetic flux label $\rho$, 
which gives the distance to the last-closed flux surface (LCFS) as mapped to the outboard
mid-plane along magnetic field lines. This coordinate is calculated by magnetic
equilibrium reconstruction with the EFIT code using input from
a set of magnetic diagnostics installed in the vacuum vessel \Ref{granetz-1990}. 
For positions in the near and far scrape-off layer, the probe was targeted to dwell 
at $\rho \approx 3\, \mathrm{mm}$ and at $\rho \approx 8\, \mathrm{mm}$ respectively. 
The north-east (NE) and south-east (SE) electrodes were biased to $-290\,\mathrm{V}$ with respect 
to the vacuum vessel in order to sample the ion saturation current. The south-west (SW) and 
north-west (NW) electrodes were electrically floating. This allows to estimate the poloidal 
electric field from these electrodes as 
\begin{align}
    E \approx \frac{V^{\mathrm{SW}} - V^{\mathrm{NW}}}{\triangle_\mathrm{p}},
\end{align}
where $\triangle_\mathrm{p} = 2.24\,\mathrm{mm}$ is the poloidal separation between 
the electrodes. 
%
 %%%%%%%%%%%%%%%%%%%%%%%%%%%%%%%%%%%%%%%%%%%%%%%%%%%%%%%%%%%%%%%%%%%%%%%%%%%%%%%%%%%%%%%%%%%%%%%
%%%%%% Vertical scanning probe
The vertical scanning probe was set up to plunge through the scrape-off layer
up to the last closed flux surface, as depicted by the vertical line in \Figref{acm_cross},
three times per plasma discharge.  A triangular voltage waveform, sweeping from $-255V$ to 
$55\, \mathrm{V}$ with
a frequency of $2\, \mathrm{kHz}$ was applied to all four electrodes of the probe head. 
The electron temperature $\Te$ is obtained by fitting a three parameter exponential function
on the measured voltage-current characteristic of each probe head with a spatial
resolution of $\triangle_{\rho} = 1\, \mathrm{mm}$ \Ref{hutch-book}.

%
%%%%%%%%%%%%%%%%%%%%%%%%%%%%%%%%%%%%%%%%%%%%%%%%%%%%%%%%%%%%%%%%%%%%%%%%%%%%%%%%%%%%%%%%%%%%%%%
%%%%%%% Description of the divertor probe arrays %%%%%%%%%%%%%%%%
The Langmuir probe array embedded in the lower divertor baffle consists of two electrodes per 
probe which were configured to sample the ion saturation current and floating potential 
respectively with $0.4\, \mathrm{MHz}$ with $16$ bit resolution.
In the targeted magnetic equilibrium configuration the two outermost divertor probes map to  
$\rho \approx 8-10\, \mathrm{mm}$. This corresponds to the approximate position 
where the horizontal scanning probe was dwelled in the far scrape-off layer within error margins 
of $5\, \mathrm{mm}$.
%
%%%%%%%%%%%%%%%%%%%%%%%%%%%%%%%%%%%%%%%%%%%%%%%%%%%%%%%%%%%%%%%%%%%%%%%%%%%%%%%%%%%%%%%%%%%%%%%

% We report from measurements obtained in 7 ohmically heated, lower single null discharges.
% For five of the discharges, the plasma current was set to $\Ip = 0.6\,\mathrm{MA}$ and the 
% line-averages particle density was increased from $\neng = 0.15$ to $\neng = 0.42$. 
% The other two discharges feature a line-averaged particle density of $\neng \approx 0.3$ 
% with plasma currents of $\Ip = 0.8\,\mathrm{MA}$ and $\Ip=1.2\,\mathrm{MA}$ respectively. 
% The magnetic field strength is $B_0 = 5.4\,\mathrm{T}$ on the magnetic axis for all discharges.
% The magnetic geometry is fixed for discharges with constant plasma current.
% For discharges with $\Ip = 0.6\,\mathrm{MA}$ the target safety factor containing
% 95\% of the magnetic flux, $q_{95}$ was aimed to be 6.0, for $\Ip = 0.8\, \mathrm{MA}$
% $q_{95} = 4.5$ and $q_{95} = 3.5$ for $\Ip = 1.2\,\mathrm{MA}$.

\begin{figure}[htb]
    \begin{minipage}{\textwidth}
        \includegraphics[width=8.5cm]{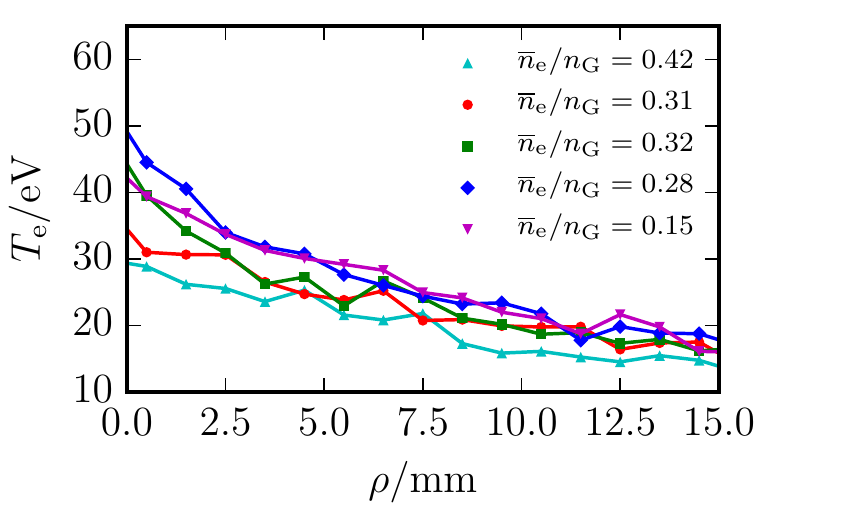}
        \caption{Radial profiles of the electron temperature as measured by the
        vertical scanning probe.}
        \label{fig:fsp_profile_te}
    \end{minipage}
\end{figure}

\begin{figure}[htb]
\begin{minipage}{\textwidth}
    \includegraphics[width=8.5cm]{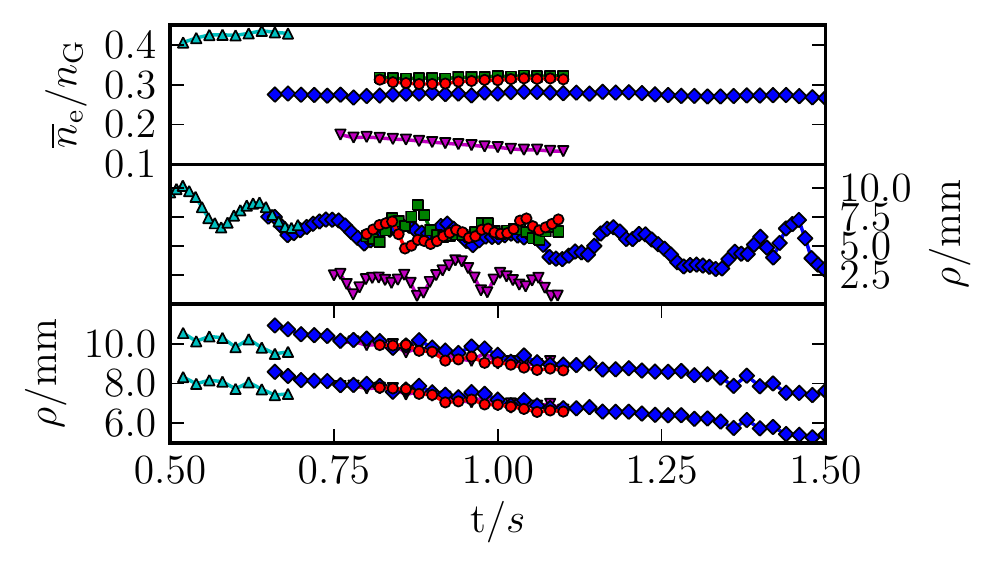}
    \caption{Evolution of the line-averaged particle density (upper panel) and
        radial coordinate for the horizontal scanning probe (mid panel). An offset of
        $\rho_0 = 5\, \mathrm{mm}$ has been added to the position of the horizontal
        scanning probe. The lower panel shows the radial coordinate for the two outer 
        most divertor probes. Table \ref{tab:shotparams} lists the used plot markers.}
    \label{fig:probes_rho}
\end{minipage}
\end{figure}

%%%% Positioning of probe head and outer divertor probe.
We report from measurements obtained in 5 ohmically heated plasmas in a lower single
null magnetic geometry with $5.4\, \mathrm{T}$ on-axis magnetic field and 
a plasma current of $\Ip = 0.6\, \mathrm{MA}$. 
For all discharges it was attempted to minimize the movement of the strike point of the
last closed flux surface on the lower divertor baffle.
As a consequence, the estimated position of the last closed flux surface at the outboard 
mid-plane is subject to larger fluctuations.
Table \ref{tab:shotparams} lists the plasma parameters of all shots as well as the position of 
the horizontal scanning probe, the time interval on which the time series are 
analyzed, and the plot marker used in the following figures. The electron temperature
at $\rho = 5 \mathrm{mm}$ which is used to normalize the floating potential and to estimate
the acoustic velocity at the position of the horizontal scanning probe is also listed.
In discharge $1$ the horizontal scanning probe was dwelled in the near scrape-off layer, 
this data is not directly comparable to the far scrape-off layer data. In discharge 3 
the sensitivity of the electronics of the divertor probes was not adjusted correctly 
such that this data is not analyzed either. Radial profiles of the electron temperature 
are shown in \Figref{fsp_profile_te}. 

The upper panel of \Figref{probes_rho} shows the time traces of the line-averaged particle 
density for the analyzed discharges. The middle panel shows the radial coordinate 
of the probe head of the horizontal scanning probe and the lower panel shows the 
radial coordinates of the two outermost divertor probes. While the line-averaged plasma density
is approximately constant and the radial coordinate of the divertor probes show a slight drift,
the radial coordinate of the horizontal scanning probe is subject to larger fluctuations.
The indicated time intervals in this figure correspond to the interval of the
time series used for data analysis. These time intervals are chosen such as to keep the 
line-averaged particle density of any given discharge within $\triangle_{\neng} \approx 0.02$
and the radial position of the horizontal scanning probe within an interval of 
$\triangle_\rho \approx 5 \mathrm{mm}$ of the reference position.

\begin{table}[h!tb]
\begin{tabular}{l| c| c| c| c| c| c|  c|}
    Discharge & $\neng$ & $\Te / \mathrm{eV}$ &  Probe position   & $t_{\mathrm{start}} / \text{s}$  & $t_{\mathrm{end}} / \text{s}$ & Plot marker\\ \hline
    1         & $0.15$  & $35$                &  near SOL       & $0.75$ ($0.75$)                   & $1.10$ ($1.10$)                & \includegraphics[height=2ex]{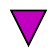} \\
    2         & $0.28$  & $25$                &  far SOL        & $0.65$ ($0.65$)                   & $1.50$ ($1.50$)                & \includegraphics[height=2ex]{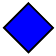} \\
    3         & $0.32$  & $25$                &  far SOL        & $0.80$ (--)                       & $1.10$ (--)                    & \includegraphics[height=2ex]{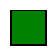} \\ 
    4         & $0.31$  & $20$                &  far SOL        & $0.80$ ($0.80$)                   & $1.10$ ($1.10$)                & \includegraphics[height=2ex]{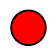} \\
    5         & $0.42$  & $20$                &  far SOL        & $0.50$ ($0.50$)                   & $0.70$ ($0.70$)                & \includegraphics[height=2ex]{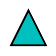} \\
\end{tabular}
\caption{List of the plasma parameters and the time interval used for time series analysis.
    The  numbers in parenthesis give the interval on which data from the divertor probe is used. 
    A dash indicates that no data is available.}
\label{tab:shotparams}
\end{table}
%
%
%%%%%%%%%%%%%%%%%%%%%%%%%%%%%%%%%%%%%%%%%%%%%%%%%%%%%%%%%%%%%%%%%%%%%%%%%%%%%%%%%%%%%%%%%%%%%%%%%%%
%%%%%%%%%%%%%%%%%%%%%%%%%%%%%%%%%%%%%%%%%%%%%%%%%%%%%%%%%%%%%%%%%%%%%%%%%%%%%%%%%%%%%%%%%%%%%%%%%%%
%%%%%%%%%%%%%%%%%%%%%%%%%%%%%%%%%%%%%%%%%%%%%%%%%%%%%%%%%%%%%%%%%%%%%%%%%%%%%%%%%%%%%%%%%%%%%%%%%%%
%
% Analysis of the outboard mid-plane fluctuations
%
% * Figure 5: Analysis of histogram shape, I_rms / <I>, S, F: shot 2
% * Figure 6: Analysis of histogram shape, I_rms / <I>, S, F: shot 5
% * Presentation of fit parameters, discussion of the goodness of fit on the histograms
% * Intermittency analysis: tau_wait and amplitude distribution
%
%

\section{Outboard mid-plane plasma fluctuations}
\label{sec:sol_results}

\begin{figure}[htb]
\begin{minipage}{\textwidth}
    \includegraphics[width=8.5cm]{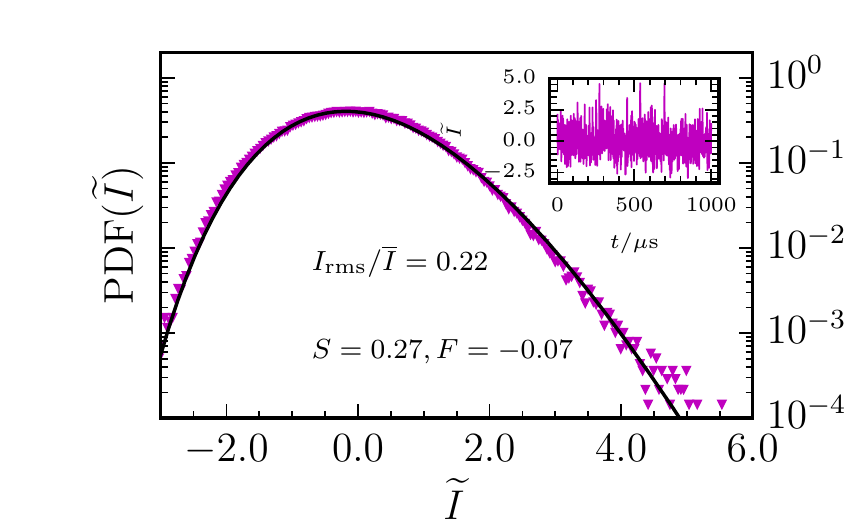}
    \caption{Histogram of the normalized ion saturation current as sampled by the north-east electrode 
        of the horizontal scanning probe dwelled in the near scrape-off layer during discharge 1 with $\neng = 0.15$. 
        The black line indicates the best fit of the stochastic model on the histogram and the inset shows a $1 \mathrm{ms}$
        long sub record of the normalized ion saturation current time series.}
    \label{fig:asp_histfit_007}
\end{minipage} 
\end{figure}

%%%%%%%%%%%%%%%%%%%%%%%%%%%%%%%%%%%%%%%%%%%%%%%%%%%%%%%%%%%%%%%%%%%%%%%%%%%%%%%%%%%%%%%%%%%%%%%%%%%
%
%%%%%%%%%%%%% Part 1) Discuss the histograms for shots 1, 2 and 5
%
%%%%%%%%%%%%%%%%%%%%%%%%%%%%%%%%%%%%%%%%%%%%%%%%%%%%%%%%%%%%%%%%%%%%%%%%%%%%%%%%%%%%%%%%%%%%%%%%%%%
We begin by analyzing the time series sampled by the horizontal scanning probe in the
near scrape-off layer. This is discharge $1$ in \tabref{shotparams} with $\neng = 0.15$.
The histogram of the normalized time series, shown in \Figref{asp_histfit_007}, is almost
gaussian and the time series, shown in the inset of the figure, appears to be random.
With a mean value of $\smean{I} = 6.1 \times 10^{-2} \mathrm{A}$ and a root mean square value
given by $\rms{I} = 1.4 \times 10^{-2}$ the relative fluctuation level of the time series
is $\rms{I} / \smean{I} = 0.22$. Coefficients of skewness and excess kurtosis are given by
$S = 0.27$ and $F = -0.07$ respectively. A best fit on the model \Eqnref{noisy_sn} yields
$\gamma = 30$ and $\epsilon = 6.24 \times 10^{-6}$. This describes a process with mostly
gaussian statistics, as suggested by the histogram and the statistics of the time series.

\begin{figure}[htb]
\begin{minipage}{\textwidth}
    \includegraphics[width=8.5cm]{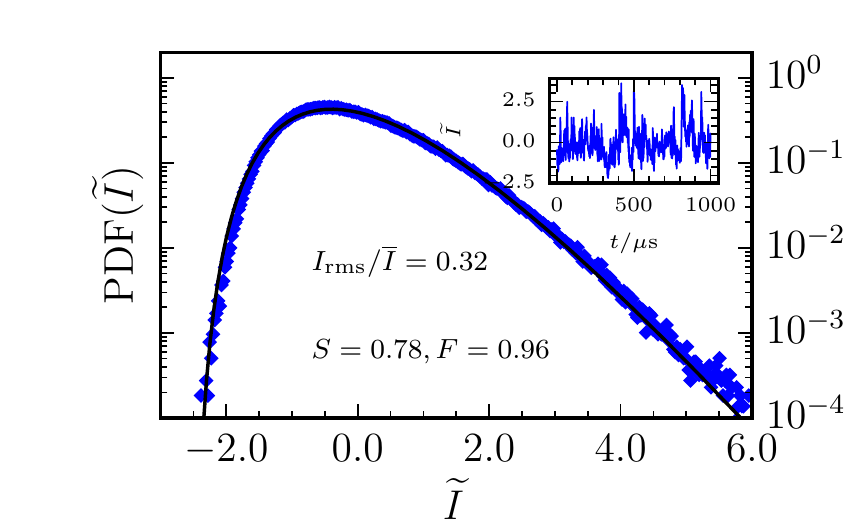}
    \caption{Histogram of the normalized ion saturation current as sampled by the north-east electrode 
        of the horizontal scanning probe dwelled in the far scrape-off layer during discharge 2 with $\neng = 0.28$. 
        The black line indicates the best fit of the stochastic model on the histogram and the inset shows a $1 \mathrm{ms}$
        long sub record of the normalized ion saturation current time series.}
    \label{fig:asp_histfit_008}
\end{minipage} 
\end{figure}

We continue by analyzing the time series sampled by the horizontal scanning probe in the 
far scrape-off layer. 
Figure \ref{fig:asp_histfit_008} shows the histogram of the ion saturation current, normalized
according to \Eqnref{is_normalization}, as sampled by the north-east electrode of the horizontal 
scanning probe during discharge $2$ with $\neng = 0.28$. The length of the time series is 
$0.85\,\mathrm{s}$ and its histogram spans over four decades in normalized probability. It 
presents an elevated tail with fluctuations exceeding six times the root mean square of 
the time series. The raw time series prominently features positive, large amplitude 
bursts events. The histogram of data sampled by the south-east electrode is quantitatively 
similar. A sample mean of $\smean{\Isat} = 4.0 \times10^{-2} \, \mathrm{A}$ and 
$\rms{\Isat} = 1.3 \times10^{-2} \,\mathrm{A}$, 
yields a normalized fluctuation level of $\rms{\Isat} / \smean{\Isat} = 0.32$. Sample 
coefficients of skewness and excess kurtosis are given by $S = 0.78$ and $F = 0.96$. A 
non-linear least squares fit on the model described by \Eqnref{noisy_sn} yields 
$\gamma = 6.35$ and $\epsilon = 2.6 \times 10^{-5}$. This describes the situation where 
the fluctuation level of the background fluctuations is well below the 
fluctuation level introduced by the bursts in the time series.

\begin{figure}
\begin{minipage}{\textwidth}
    \includegraphics[width=8.5cm]{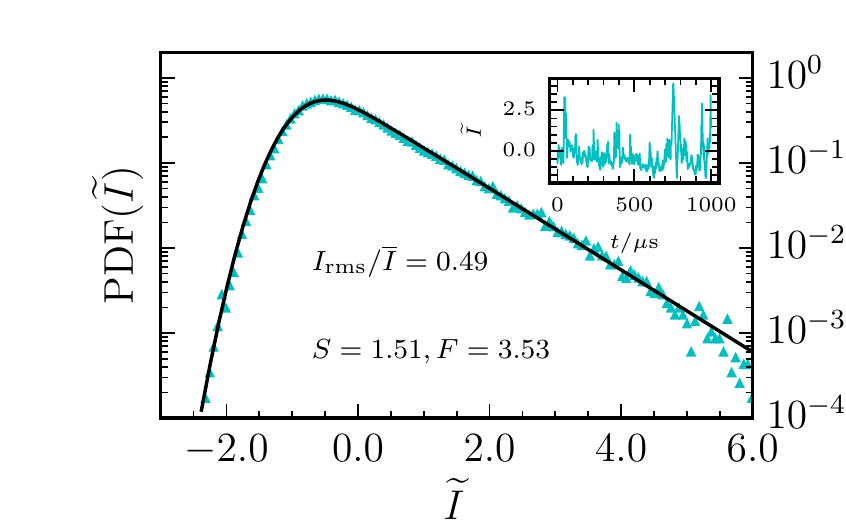}
    \caption{Histogram of the normalized ion saturation current as sampled by the north-east electrode 
        of the horizontal scanning probe dwelled in the far scrape-off layer during discharge 5 with $\neng = 0.42$. 
        The black line indicates the best fit of the stochastic model on the histogram and the inset shows a $1 \mathrm{ms}$
        long sub record of the normalized ion saturation current time series.}
        \label{fig:asp_histfit_012}
\end{minipage}
\end{figure}

Figure \ref{fig:asp_histfit_012} shows the histogram of the normalized ion saturation 
current time series sampled during discharge $5$ with $\neng = 0.42$. The histogram 
presents an elevated tail with fluctuations well exceeding six times the root mean square 
of the time series. The mean of the time series is given by 
$\smean{\Isat} = 9.4 \times 10^{-2} \mathrm{A}$ and its root mean square value is given 
by $\rms{\Isat} = 4.6 \times 10^{-2} \mathrm{A}$. This yields a normalized fluctuation 
level of $\rms{\Isat} / \smean{\Isat} = 0.49$, coefficients of skewness and excess 
kurtosis are given by $S=1.5$ and $F=3.5$. The time series presents positive, large 
amplitude burst events, which seem to occur less frequent as in \Figref{asp_histfit_008}. 
The best fit on the model given by \Eqnref{noisy_sn} yields $\gamma = 1.06$ and 
$\epsilon = 1.85 \times 10^{-1}$, suggesting that background fluctuations are of larger 
relative magnitude than in the previous case.

%
%%%%%%%%%%%%%%%%%%%%%%%%%%%%%%%%%%%%%%%%%%%%%%%%%%%%%%%%%%%%%%%%%%%%%%%%%%%%%%%%%%%%%%%%%%%%%%%%%%%
%
%%%%%%%%%%%%% Part 2) Continue by fitting the statistical models on the data
%
%%%%%%%%%%%%%%%%%%%%%%%%%%%%%%%%%%%%%%%%%%%%%%%%%%%%%%%%%%%%%%%%%%%%%%%%%%%%%%%%%%%%%%%%%%%%%%%%%%%

\begin{figure}[htb]
    \includegraphics[width=8.5cm]{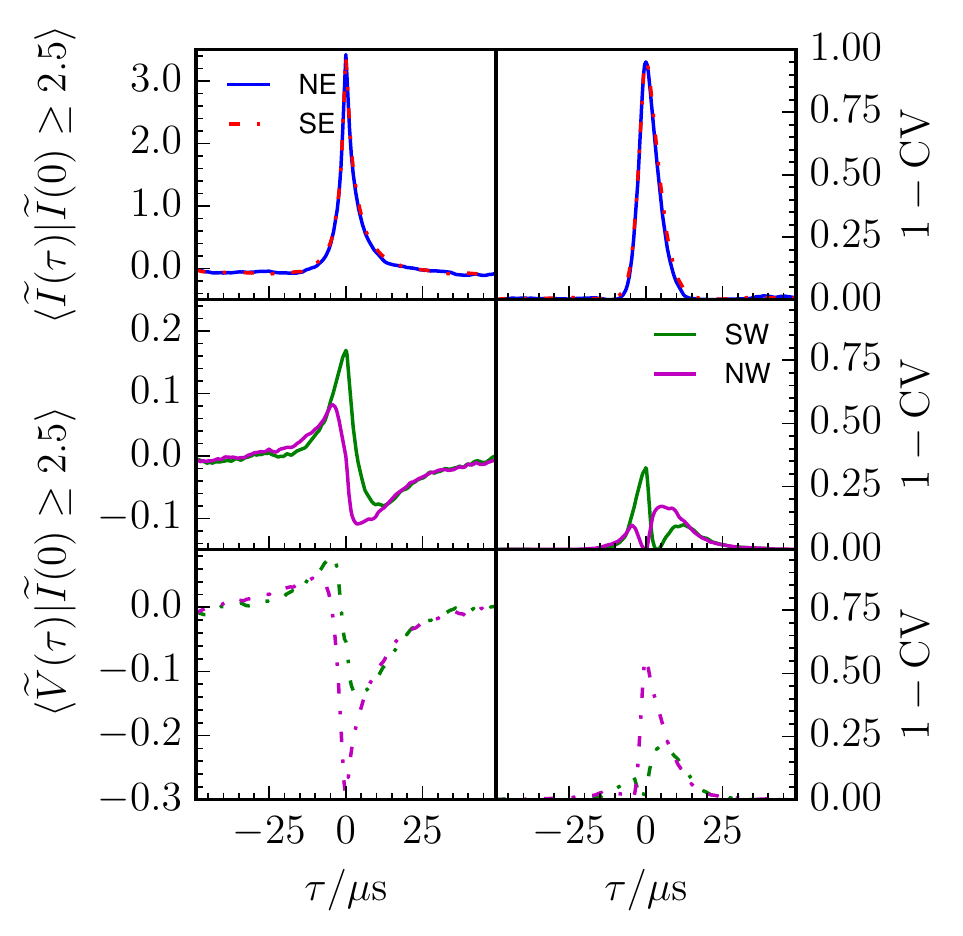}
    \caption{Conditionally averaged burst shape and conditional variance for the ion saturation current (upper row), 
    floating potential when triggered by bursts on the north-east electrode (middle row), and floating 
    potential when triggered by bursts on the south-east electrode (bottom row) for discharge 2 with $\neng = 0.28$.}
    \label{fig:asp_condavg_008}
\end{figure}

Figure \ref{fig:asp_condavg_008} shows the conditionally averaged waveforms and their
conditional variance of the normalized data time series sampled during discharge $2$ with
$\neng = 0.28$. The upper row shows the conditionally averaged waveform of large-amplitude
bursts occurring in the ion saturation current, as measured by the north-east and south-east
electrodes, as well as their conditional variance. 
The averaged waveform is asymmetric. The best fit of an exponential waveform on the rise
and fall give an e-folding rise time of $\taur \approx 2\, \mu \mathrm{s}$ and fall time 
of $\tauf \approx 4\, \mu \mathrm{s}$ respectively.
Their reproducibility is close to unity within the interval centered around 
$\tau=0\, \mu \mathrm{s}$, bounded by the e-folding times, and it shows the same asymmetry
as the burst shape. 

The conditionally averaged floating potential waveform, computed by setting the
trigger condition on bursts in the ion saturation current time series as sampled by the
north-east electrode, is shown in the middle row of \Figref{asp_condavg_008}.
The south-west electrode measures a dipolar waveform where the positive
peak is sampled before the negative peak. The peak-to-valley range
of the waveform is approximately 
%$7\, \mathrm{V}$ 
$0.3$ % using T_e = 25eV
where the positive peak is larger in absolute value than the negative peak by a factor of $2$.
The waveform sampled by the north-west electrode is more symmetric, and features a peak-to-valley 
range of approximately $0.2$. The positive peak is also more reproducible with
$1 - \mathrm{CV} \approx 0.3$ compared to $1 - \mathrm{CV} \approx 0.2$ for the
north-west electrode.

Rather triggering on the south-east electrode, the conditionally averaged floating potential waveforms 
are also dipolar with peak-to-valley ranges of approximately
$0.2\, (0.4)$ %using T_e = 25eV
for the south-west (north-west) electrode.
The reproducibility of the waveform is larger by a factor of two for the
latter. Opposite to the situation where the trigger is on the north-east electrode,
here the reproducibility is larger on the north-west electrode where the negative 
part of the blobs electric potential dipole is measured after its density maximum
has traversed the probe. 

Such poloidal motion is in agreement with the picture of a dipolar electric potential 
structure, centered around the particle density maximum of a plasma blob which is 
moving into the direction of $\bm{B} \times \nabla B$, i.e. poloidally downwards. This 
is compatible with measurements using gas-puff imaging 
\Ref{zweben-2006, agostini-2011}. 
For the plasma blob to propagate radially outwards, the negative pole of the electric 
potential has to be poloidally above the particle density maximum and the positive pole 
has to be poloidally below the particle density maximum. When the particle density 
maximum is recorded by the north-east electrode, the positive pole of the potential 
structure has traversed the south-west electrode. This explains the pronounced positive 
pole for $\tau < 0$ of the south-west electrode and its relatively large reproducibility.
The negative pole of the potential structure traverses the north-west electrode for 
$\tau > 0$ and leads to a large reproducibility of the waveform.

The conditionally averaged waveform of the estimated poloidal electric field
is a monopolar structure with a peak value of approximately $-2500\, \mathrm{Vm}^{-1}(-3000\, \mathrm{Vm}^{-1})$ when triggered on bursts
occurring on the north-east (south-east) electrode. Using that the toroidal
magnetic field at the probe position is approximately $4.0\, \mathrm{T}$, this
corresponds to a local average electric drift velocity of $\vrad \approx 600 - 700\, \mathrm{ms}^{-1}$. 
Radial blob velocities of similar magnitude have been reported from gas-puff imaging 
measurements \Ref{agostini-2011, kube-2013}.

We continue by elucidating the relation between the amplitudes of the bursts and their 
associated radial velocity. For this, we approximate the time it takes for a blob to 
traverse the probe by $\taur + \tauf$. Both e-folding times are found by a least squares 
fit of an exponential function on the rise and fall of the conditionally averaged burst 
shape respectively. The electric drift velocity associated with a burst event is then 
computed using the estimated poloidal electric field averaged over the interval 
$[ -\taur : \tauf]$. 

\begin{figure}[htb]
    \includegraphics[width=8.5cm]{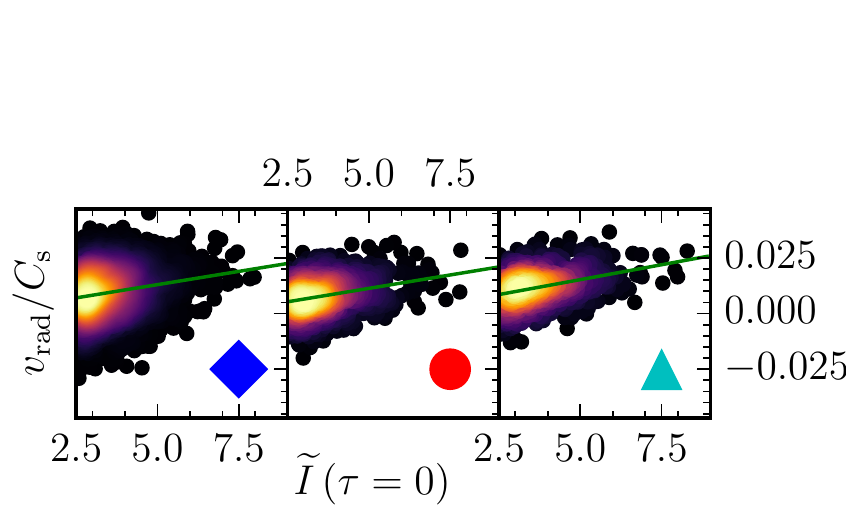}
    \caption{Estimated radial velocity as a function of normalized burst amplitude with
        the best fit of a linear function denoted by the green line. The data is sampled by the horizontal
        scanning probe dwelling in the far scrape-off layer and the plot marker refers to \Tabref{shotparams}.}
        \label{fig:scatter_epol_isat}
\end{figure}

Figure \ref{fig:scatter_epol_isat} shows the radial electric drift velocity associated with 
burst events in the ion saturation current on the north east electrode plotted against 
its normalized amplitude. The radial velocities rarely exceed $5\%$ of the ion acoustic 
velocity. Approximately $90\%$ of all events have a velocity directed towards the main 
chamber wall and the Pearson sample correlation coefficient increases from $r = 0.19$ for 
$\neng = 0.28$ (left panel, 5253 events), to $r = 0.29$ for $\neng = 0.31$ (middle panel, 
1203 events), to $r = 0.36$ for $\neng = 0.42$ (right panel, 833 events). To guide the 
eye on the sample correlation, a green straight line, denoting a linear fit on the value 
pairs, has been over plotted in all scatter plots.

%
%%%%%%%%%%%%%%%%%%%%%%%%%%%%%%%%%%%%%%%%%%%%%%%%%%%%%%%%%%%%%%%%%%%%%%%%%%%%%%%%%%%%%%%%%%%%%%%%%%%%%%%%%%%%%%
Conditional averaging further reveals the distribution of waiting times between
successive large amplitude burst events and of the burst amplitudes of the normalized time
series at hand. For discharges where multiple electrodes sample the ion saturation current,
only data sampled by the north east electrode is used.

The shape of the sampled histograms indicates that the waiting times and the burst 
amplitudes are approximately described by an exponential distribution. The description
by the exponential model appears more accurate for the waiting times than for the
burst amplitudes.
For an exponentially distributed random variable $X > 0$, the complementary 
cumulative distribution function is given by
\begin{align}
    1 - F_X\left( X \right) & = \bigexp{-\frac{ X - X_0 }{\mean{X}}} \label{eq:PDF_expon}.
\end{align}
Here $F_X$ is the cumulative distribution function, $\mean{X}$ is the scale parameter of the 
distribution, in the cases discussed here the average waiting time and average burst 
amplitude, and $X_0$ is the location parameter of the distribution. To obtain the average 
waiting time of the distribution from sampled data we employ a maximum likelihood 
estimate. This method is unbiased in the sense that all data points are equally weighted 
when estimating the scale parameter \Ref{fraile-2005}. The location parameter is given by 
the conditional averaging sub record length in the case of waiting time distributions and 
the conditional averaging threshold in the case of burst amplitude distributions.

\begin{figure}[htb]
\begin{minipage}{\textwidth}
    \includegraphics[width=8.5cm]{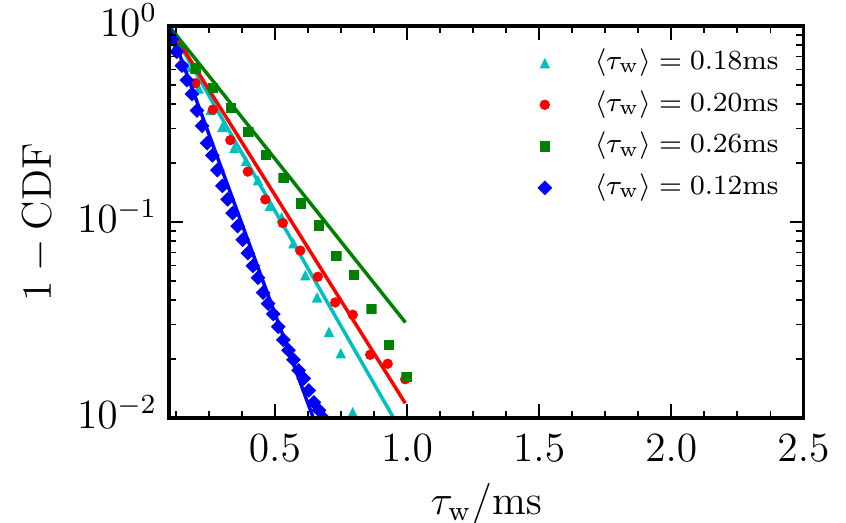}
    \caption{Distribution of waiting times between successive large amplitude burst events
        in the normalized ion saturation current time series as measured by the horizontal scanning probe
        in the far scrape-off layer. The plot markers refer to discharges listed in \Tabref{shotparams}.}
    \label{fig:asp_tauwait}
\end{minipage}
\end{figure}

Figure \ref{fig:asp_tauwait} shows the histograms of the sampled waiting times between 
successive burst events with amplitudes exceeding $2.5$. Compared are best fits on 
\Eqnref{PDF_expon}, denoted by full lines. The exponential model gives a good description
of the waiting times over more than one decade in normalized probability for all discharges.
The average waiting time is between
$\mean{ \tauw} \approx 0.12\, \mathrm{ms}$ for $\neng = 0.28$, 
$0.20\, \mathrm{ms}$ and $0.26\, \mathrm{ms}$ for discharges 3 and 4,
and $0.18\, \mathrm{ms}$ for discharge 5 where $\neng = 0.42$.
No trend between the line-averaged plasma density and the average waiting time is observed.

Histograms of the sampled normalized burst amplitudes are compared to the best fits of
\Eqnref{PDF_expon} in \Figref{asp_amean}. We find that the burst amplitude
histograms are approximately described by an exponential distribution over approximately
one decade. However, as all time series feature significant pulse overlap, the burst 
amplitude histogram is only suggestive of the actual amplitude distribution of the 
individual pulses that make up the signal. The average burst value is between
$\mean{A} = 1.1$ for $\neng = 0.28$ and  $\mean{A} = 1.3$ for $\neng = 0.42$, with no
apparent correlation to the line-averaged density. That is, the average burst amplitude 
is approximately given by the root mean square value of the time series.

%%%%%%%%%%%%%%%%%%%%%%%%%%%%%%%%%%%%%%%%%%%%%%%%%%%%%%%%%%%%%%%%%%%%%%%%%%%%%%%%%%%%%%%%%%%%%%%%%%%%%%%%%%%%%%%%%%%%%%%%
%
% Divertor plasma fluctuations
%
%%%%%%%%%%%%%%%%%%%%%%%%%%%%%%%%%%%%%%%%%%%%%%%%%%%%%%%%%%%%%%%%%%%%%%%%%%%%%%%%%%%%%%%%%%%%%%%%%%%%%%%%%%%%%%%%%%%%%%%%

\section{Divertor plasma fluctuations}
\label{sec:div_fluct}
We proceed by analyzing data time series sampled by the Langmuir probes embedded in the lower 
divertor in the same manner as in the previous section. 
Figure \ref{fig:osd_histfit_007} presents the histogram of the normalized ion saturation 
current time series for discharge $1$ with $\neng = 0.15$, as sampled by the two outermost 
divertor probes 9 (shown in the upper panel) and 10 (shown in the lower panel). 
The average current at probe 9 is 
$\smean{\Isat} = 4.4 \times 10^{-2} \mathrm{A}$
and the root mean square value is given by
$\rms{\Isat} = 1.5 \times 10^{-2} \mathrm{A}$, 
which yields a relative fluctuation level of 
$\rms{\Isat} / \smean{\Isat} = 0.33$. 
As shown in the upper panel of \Figref{osd_histfit_007}, the sample presents only slightly
elevated tails, fluctuations in the time series rarely exceed four times the root mean square 
value of the time series. Coefficients of skewness and excess kurtosis are given by
$S = 0.41$ and $F = -0.28$. 
The histogram of the ion saturation current as sampled by divertor probe 10
presents a more elevated tail with fluctuations exceeding five times the root mean 
square value of the time series. With 
$\smean{\Isat} = 2.9 \times 10^{-2}\mathrm{A}$ 
and
$\rms{\Isat} = 9.0 \times 10^{-3} \mathrm{A}$ 
the relative fluctuation level is 
$\rms{\Isat} / \smean{\Isat} = 0.31$.
%%%%%%%%%%%%%%%%%%%%%%%%%%%%%%%%%%%%%%%%%%%%%%%%%%%%%%%%%%%%%%%%%%%%%%%%%%%%%%%%
The best fit on the model \Eqnref{noisy_sn} yields $\gamma = 9.9\, (5.8)$ and 
$\epsilon = 2.0 \times 10^{-5}\, (7.3 \times 10^{-4})$ for the time series sampled by
probe $9\, (10)$.

Figure \ref{fig:osd_histfit_012} presents the histograms of the normalized ion saturation 
current time series sampled by the divertor probes for discharge $5$ with $\neng = 0.42$. 
Both time series present fluctuations of up to five times the sample root mean square value.
For the time series obtained by probe $9$ the sample mean is given by
$\smean{\Isat} = 0.20\, \mathrm{A}$ and the root mean square value is
given by $\rms{\Isat} = 7.8 \times 10^{-2}\, \mathrm{A}$. This gives a normalized 
fluctuation level of $\rms{\Isat} / \smean{\Isat} = 0.38$. 
Sample coefficients of skewness and excess kurtosis are given by $S = 1.3$ and $F = 2.3$,
which reflects the non-gaussian character of the fluctuations. 
Continuing with the histogram of the normalized ion saturation current time series sampled 
by probe 10, shown in the lower panel of \Figref{osd_histfit_012}, we find its histogram 
to be similar to the histogram sampled by probe 9. The ion saturation current fluctuation 
amplitudes do not exceed six times the sample root mean square value. Values of the sample 
mean, root-mean square and relative fluctuation level are given by
$\smean{\Isat} = 0.20 \mathrm{A}$, 
$\rms{\Isat} = 5.7 \times 10^{-2} \mathrm{A}$, and
$\rms{\Isat} / \smean{\Isat} = 0.28$,
coefficients of sample skewness and excess kurtosis are given by $S=1.0$ and
$F=1.8$. The best fit on the model described by \Eqnref{noisy_sn} yields 
$\gamma = 2.2\, (1.4)$ and $\epsilon = 5.3 \times 10^{-2}\, (3.5 \times 10^{-1})$ for the 
time series sampled by probe 9(10).

%
%%%%%%%%%%%%%%%%%%%%%%%%%%%%%%%%%%%%%%%%%%%%%%%%%%%%%%%%%%%%%%%%%%%%%%%%%%%%%%%%%%%%%%%%%%%%%%%%%%%%%%%%%%%%%%%%%%%%%%%%%%%%%%%%%%%%%%%%%%%%%%%%%%%%%%%
%
%
%
%%%%%%%%%%%%%%%%%%%%%%%%%%%%%%%%%%%%%%%%%%%%%%%%%%%%%%%%%%%%%%%%%%%%%%%%%%%%%%%%%%%%%%%%%%%%%%%%%%%%%%%%%%%%%%%%%%%%%%%%%%%%%%%%%%%%%%%%%%%%%%%%%%%%%%%
We continue by analyzing the conditionally averaged waveforms of the time series 
sampled by probe $10$. For discharge $5$ with $\neng = 0.42$ we assume a detached 
divertor and use half the electron temperature measured by the vertical scanning 
probe, $\Te = 10\,\mathrm{eV}$, to normalize the floating potential time series \Ref{labombard-1995}. 
\Figref{osd_condavg} shows the conditionally averaged waveforms for discharges 
$1$ ($\neng = 0.15$), $2$ ($\neng = 0.28$), and $5$ ($\neng = 0.42$). 
%%%%
%
For discharges $1$ and $2$ the conditionally averaged burst shape is nearly symmetric. 
Least squares fits of an exponential function on the burst shape yield e-folding times 
of $\taur \approx 12\, \musec$ and $\tauf \approx 14\, \musec$ and 
$\taur \approx 14\, \musec$ and $\tauf \approx 12\, \musec$ respectively. The 
conditionally averaged burst shape for discharge $5$ is asymmetric with a rise time of 
$\taur \approx 26\, \musec$ and a fall time of $\tauf \approx 66\musec$. All 
conditionally averaged burst shapes are highly reproducible. 

The conditionally averaged waveform of the floating potential is shown in the lower panel
of \Figref{osd_condavg}. For discharges $1$ and $2$ the conditionally averaged floating 
potential waveforms associated with large amplitude bursts in the ion saturation current 
have a dipolar shape with a pronounced positive peak and are reproducible. For discharge 
$5$ the conditionally averaged waveform is irregular, showing only a weak positive peak, 
and is irreproducible. 
%%%%

We continue by studying the intermittency of large amplitude burst events in the 
normalized ion saturation current time series sampled by divertor probe 10. Figure 
\ref{fig:osd_tauwait} shows histograms of the waiting times between successive large 
amplitude burst events in the time series. Full lines denote \Eqnref{PDF_expon} with 
an average waiting time obtained by a maximum likelihood estimate and a location 
parameter given by $\tau_{\mathrm{w},0} = 0.1\, \mathrm{ms}$.
All histograms are well approximated by an exponential distribution over one decade in 
probability. Average waiting times between large amplitude burst events are between 
$0.28\, \mathrm{ms}$ and $0.43\, \mathrm{ms}$, approximately twice as large as observed in time 
series sampled in the outboard mid-plane scrape-off layer.
Figure \ref{fig:osd_amean} shows the histogram of the burst amplitudes in the time series.
Maximum likelihood estimates of the average burst amplitude are $\mean{A} \approx 0.6$ for 
$\neng = 0.15$ and $0.30$, which increases to $\mean{A} \approx 0.9$ for $\neng = 0.42$. 
As in the case of the horizontal scanning probe data, no systematic variation of the 
scale length with line-averaged particle density is observable. The average burst 
amplitude is approximately half the amplitude found for the time series sampled in the 
outboard mid plane scrape-off layer.

\section{Discussion}
\label{sec:discussion}
%%% New text:
Long ion saturation current time series, with sample lengths between $0.2$ and 
$0.85\, \mathrm{s}$ have been sampled in the outboard mid-plane scrape-off layer and 
at the outer divertorr, during discharges with line averaged plasma densities ranging
from $\neng = 0.15$ to $\neng = 0.42$. A statistical analysis shows that the time series 
in the far scrape-off layer are characterized by large relative fluctuation levels 
and intermittent large-amplitude burst events. The sample statistics of all ion
saturation current time series discussed in this paper are collected in 
\Tabref{ts_statistics_upstream_downstream}. The data sampled in the near scrape-off 
layer presents Gaussian statistics, consistent with previous measurements in the near 
scrape-off layer \Ref{labombard-2001}.
In the outboard midplane far scrape-off layer we find that the relative fluctuation
level increases gradually from $0.32$ for discharge $2$ ($\neng = 0.28$) to $0.49$ in 
discharge $5$ ($\neng = 0.42$). A similar increase is found for divertor probe $9$, where
$\rms{I} / \smean{I}$ increases from $0.33$ for $\neng = 0.15$, over $0.35$ for 
$\neng = 0.28$ to $0.39$ for $\neng = 0.42$. On the other hand, the relative fluctuation 
level of the time series sampled by divertor probe $10$ shows no significant change as 
the line-averaged density is changed between discharges. Sample coefficients of skewness 
and excess kurtosis are found to increase in all time series as the line-averaged density 
increases. Thus, the intermittency level increases with the line-averaged density in 
these ohmic plasmas.

\begin{table}
    \begin{tabular}{c|c|c|c|c}
        Shot    & $\neng$   & $\rms{I} / \smean{I}$           & $S$                        & $F$  \\ \hline
        1       & $0.15$    & $0.22^{*}$    / $0.33$ / $0.33$ & $0.27^{*}$    / $0.41$ / $0.41$   & $0.07^{*}$    / $-0.28$ / $-0.29$                \\
        2       & $0.28$    & $0.32$        / $0.35$ / $0.31$ & $0.78$        / $0.71$ / $0.55$   & $0.96$        / $0.53$  / $7.5 \times 10^{-3}$   \\
        3       & $0.32$    & $0.34$        / $-$    / $-$    & $1.2$         / $-$    / $-$      & $2.2$         / $-$     / $-$                    \\
        4       & $0.31$    & $0.33$        / $0.40$ / $0.37$ & $1.1$         / $0.98$ / $0.87$   & $1.9$         / $1.3$   / $0.81$                 \\
        5       & $0.42$    & $0.49$        / $0.39$ / $0.28$ & $1.5$         / $1.3$  / $1.0$    & $3.5$         / $2.3$   / $1.8$                  \\
    \end{tabular}
    \caption{Statistics of the entire time series sampled by the horizontal scanning probe / divertor probe 9 / divertor
    probe 10. The values marked with a $^{*}$ are sampled in the near scrape-off layer and are not
    directly comparable to the other values.}
    \label{tab:ts_statistics_upstream_downstream}
\end{table}

%
% Discuss comparison to stochastic model here
%
Figure \ref{fig:asp_s_vs_f} shows the sample skewness plotted against the sample excess 
kurtosis, computed for time series sub records of $20 \mathrm{ms}$, sampled in the 
outboard midplane far scrape-off layer during discharges 2 -- 4. Both $S$ and $F$ 
increase with $\neng$. A least squares fit of the model $F = a + b S^2$ on the value 
pairs yields $a =-0.20 \pm 0.04$ and $b = 1.51 \pm 0.03$. The relation between sample 
coefficients of skewness and excess kurtosis of the time series sampled by the divertor 
probes, shown in \Figref{osd_s_vs_f}, is qualitatively similar to those from the 
horizontal scanning probe. The sample coefficients have a smaller range and notably feature 
small, feature negative values of excess kurtosis. A least squares fit on the quadratic 
model yields $a = -0.50 \pm 0.02$ and $b = 1.78 \pm 0.03$. The fit parameters have similar
magnitude as found for the outboard mid-plane far scrape-off layer. The clustering of 
the sample pairs is also similar to the clustering for the horizontal scanning probe data. 
Samples taken in low line-average density discharges present smaller coefficients than 
samples taken in high line-averaged density discharges, implying that time series from 
higher density discharges are more intermittent. 

The values of sample skewness and excess kurtosis for the outboard mid-plane time series
fall in a range between $0.0 \leq S \leq 2.0$ and $0.0 \leq F \leq 6.0$. These
ranges are considerably lower than observed for a similar analysis
of gas-puff imaging data in Alcator C-Mod \Ref{garcia-2013}. In the latter case,
the view of the diagnostics includes the area of the wall shadow, characterized
by a considerably lower plasma background density. As plasma blobs propagate into 
this region, they are registered in the intensity time series as amplitudes which are
significantly larger than the background intensity signal. This leads to
large values of sample skewness and excess kurtosis.

Histograms of time series sampled in the outboard mid-plane far scrape-off layer 
present elevated tails with fluctuations exceeding six times the root mean square value 
of the time series. The time series sampled by the divertor probes show qualitatively 
the same features, albeit with a lower normalized fluctuation magnitude. 
Figures \ref{fig:hist_is_collapsed} and \ref{fig:hist_vf_collapsed} show histograms of 
the ion saturation current and floating potential time series, normalized according to 
\Eqsref{is_normalization} and (\ref{eq:vf_normalization}), sampled in all discharges
listed in \Tabref{shotparams}. The ion saturation current histograms do not collapse
perfectly but may be separated by where they are sampled. Histograms sampled in the far 
scrape-off layer present consistently slightly more elevated tails than histograms 
sampled in the divertor region. Within each group, the highest density discharges
feature the histograms with the most elevated tails. The floating potential histograms 
are approximately normal distributed. The time series sampled in the out board mid-plane 
scrape-off layer show a slightly elevated tail compared to a normal distribution in  
positive and negative abscissa regions while the time series sampled in the divertor 
region show an elevated tail in negative abscissa regions and a lowered tail in
positive abscissa regions.

The distribution of waiting times between large-amplitude bursts in ion
saturation current time series is found to be well described by an exponential
distribution for both, time series sampled in the outboard mid plane scrape-off
layer as well and in the divertor region. This suggests that the individual large 
amplitude pulses are uncorrelated and that their occurrence is governed by a 
Poisson process. It is exactly this property for which the stochastic model of
Ref. \Ref{garcia-2012} predicts a quadratic relation between skewness and excess 
kurtosis. 

\begin{figure}[htb]
\begin{minipage}{\textwidth}
    \includegraphics[width=8.5cm]{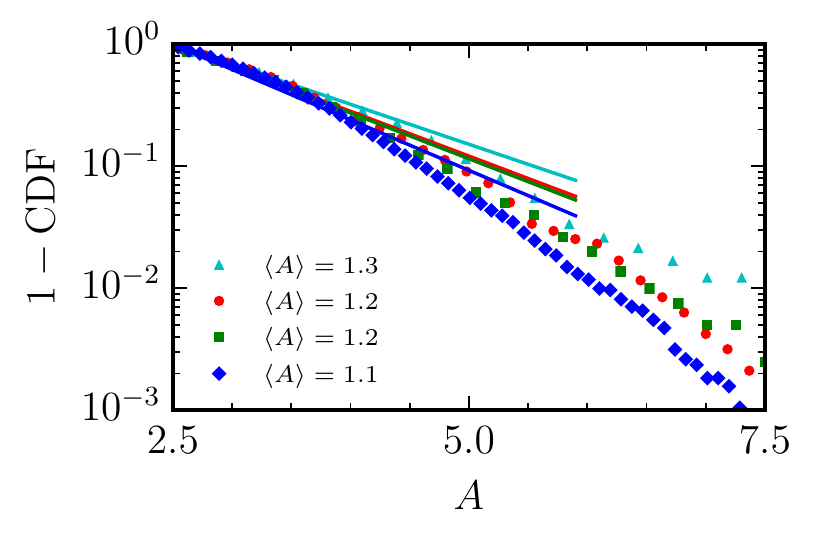}
    \caption{Amplitude distribution of burst events in the normalized ion saturation current time series 
        as measured by the horizontal scanning probe in the far scrape-off layer.
        The plot markers refer to shots as described in \Tabref{shotparams}.}
    \label{fig:asp_amean}
\end{minipage}
\end{figure}

\begin{figure}[htb]
\begin{minipage}{\textwidth}
    \includegraphics[width=8.5cm]{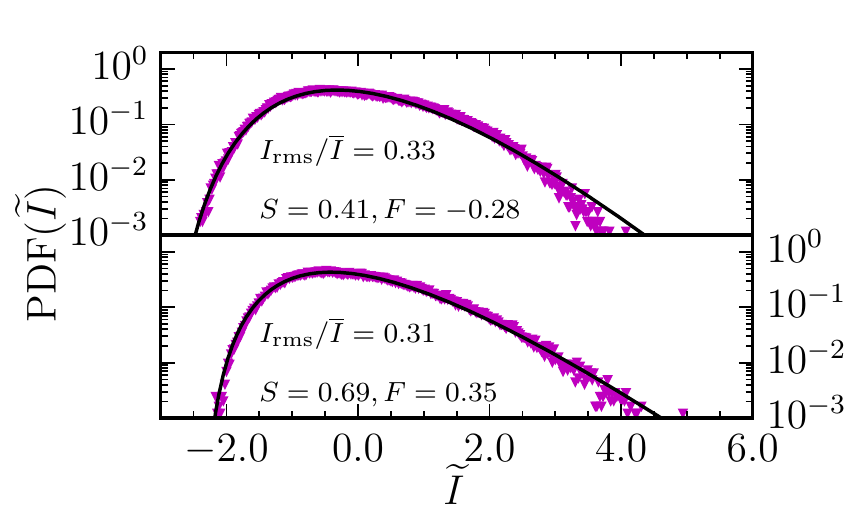}
    \caption{Histogram of the normalized ion saturation current as sampled by divertor probes 9 (upper panel),
        and 10 (lower panel) for discharge $1$ with $\neng = 0.15$. Compared is a fit on the stochastic model
        \Eqnref{noisy_sn}.}
    \label{fig:osd_histfit_007}
\end{minipage}
\end{figure}

\begin{figure}[htb]
\begin{minipage}{\textwidth}
    \includegraphics[width=8.5cm]{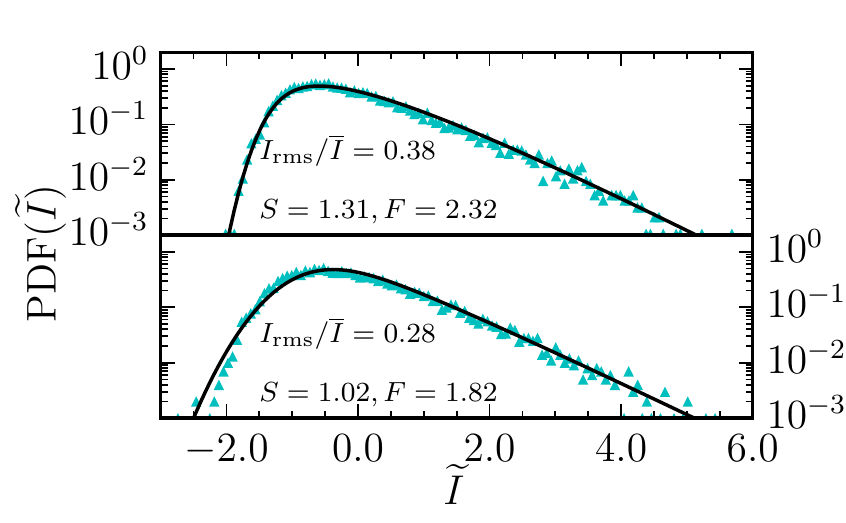}
    \caption{Histogram of the normalized ion saturation current as sampled by divertor probes 9 (upper panel),
        and 10 (lower panel) for discharge $5$ with $\neng = 0.42$. Compared is a fit on the stochastic model
        \Eqnref{noisy_sn}.}
    \label{fig:osd_histfit_012}
\end{minipage} 
\end{figure}

\begin{figure}[htb]
\begin{minipage}{\textwidth}
    \includegraphics[width=8.5cm]{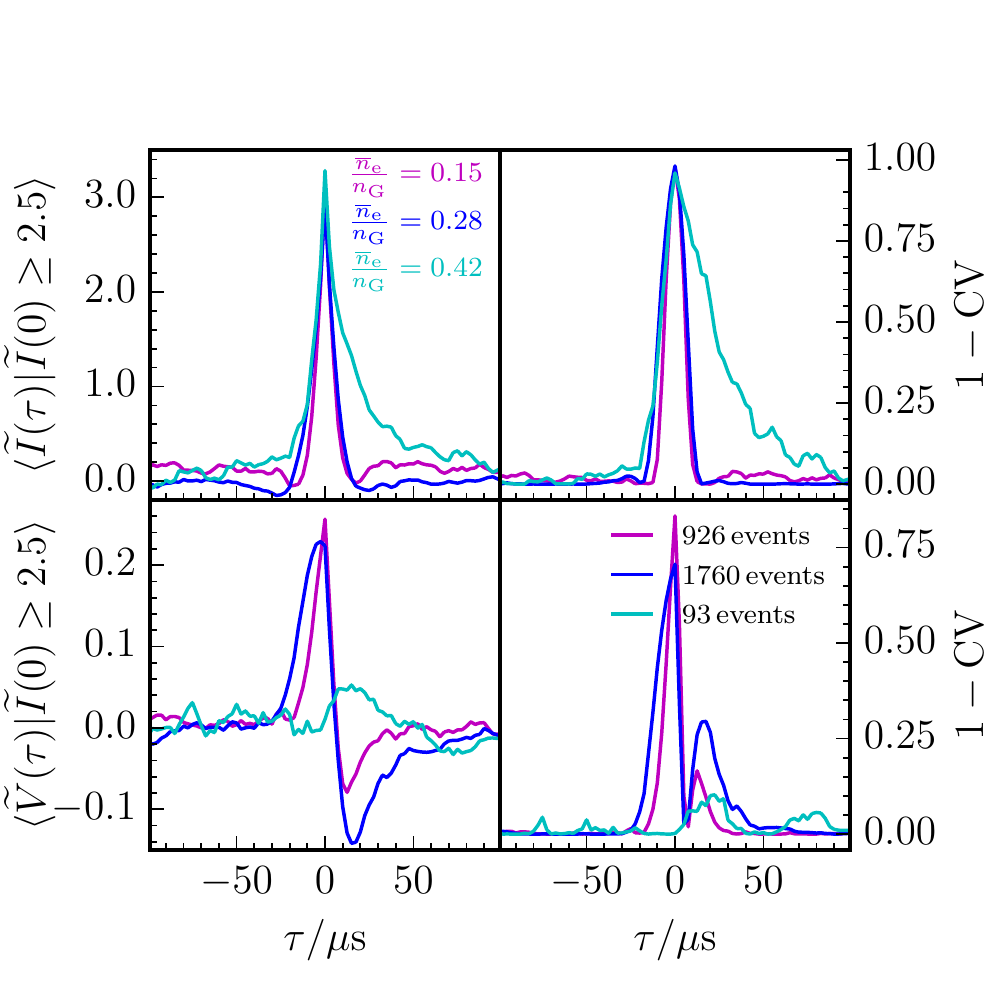}
        \caption{Conditionally averaged burst shape and conditional variance for the normalized ion saturation 
            current (upper row) and floating potential structure with conditional variance (bottom row) as 
            measured by divertor probe 10.}
    \label{fig:osd_condavg}
\end{minipage}
\end{figure}

\begin{figure}[htb]
    \begin{minipage}{\textwidth}
        \includegraphics[width=8.5cm]{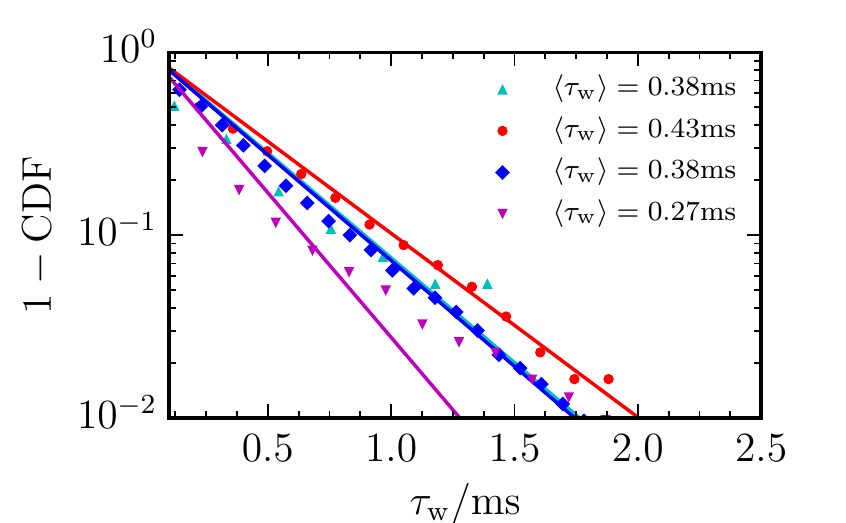}
        \caption{Distribution of waiting times between successive large amplitude burst events
            in the normalized ion saturation current time series as measured by the outermost divertor 
            probe.}
    \label{fig:osd_tauwait}
\end{minipage}
\end{figure}

\begin{figure}[htb]
    \begin{minipage}{\textwidth}
        \includegraphics[width=8.5cm]{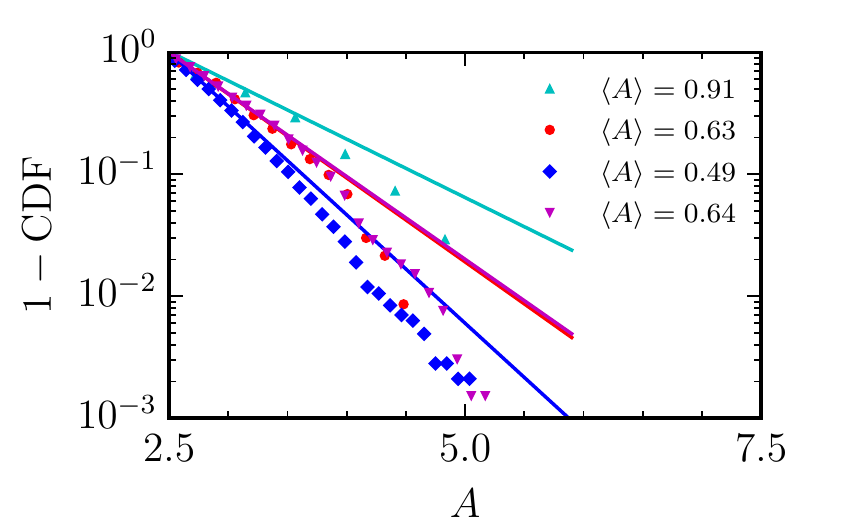}
    \caption{Amplitude distribution of burst events 
        in the ion saturation current time series as measured by the outermost divertor probe.}
    \label{fig:osd_amean}
\end{minipage}
\end{figure}

\begin{figure}
\begin{minipage}{0.5\textwidth}
    \includegraphics[width=8.5cm]{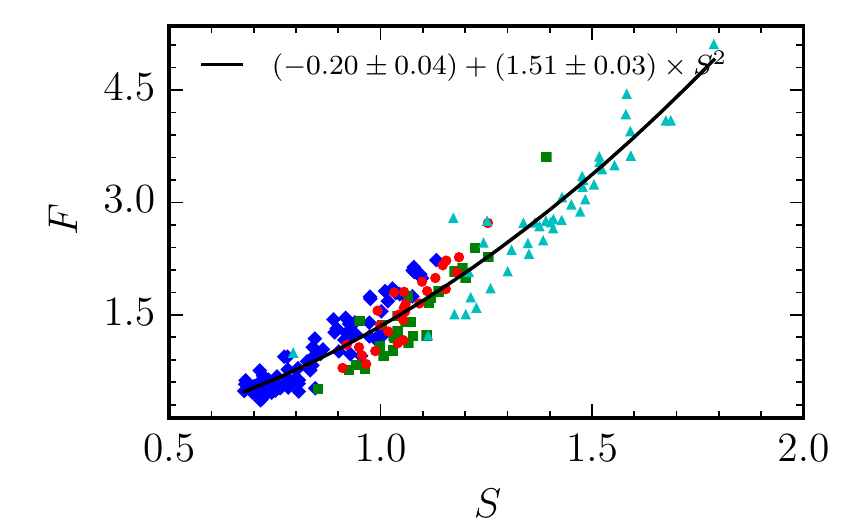}
    \caption{Coefficients of skewness and excess kurtosis computed for $20 \mathrm{ms}$ long sub 
        samples of the ion saturation current as sampled by the horizontal scanning probe.}
    \label{fig:asp_s_vs_f}
\end{minipage}
\begin{minipage}{0.5\textwidth}
    \includegraphics[width=8.5cm]{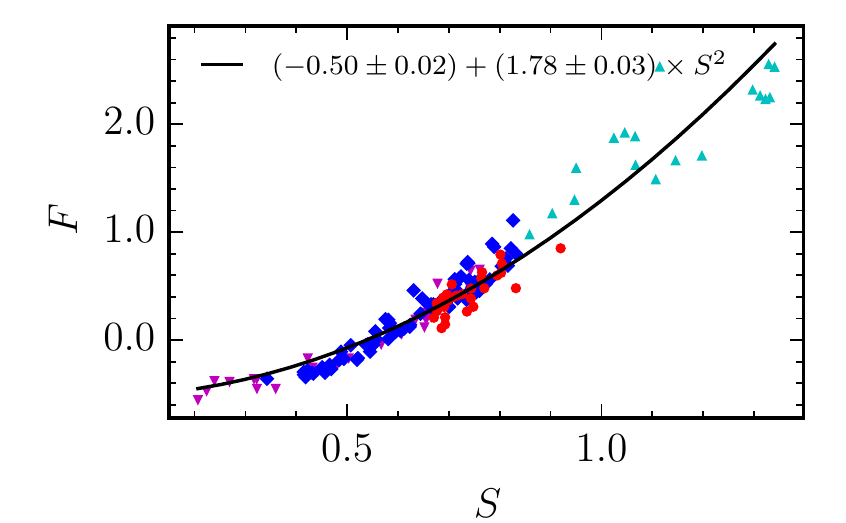}
    \caption{Coefficients of skewness and excess kurtosis computed for $20 \mathrm{ms}$ long sub 
        samples of the ion saturation current as sampled by the outermost divertor probe.}
    \label{fig:osd_s_vs_f}
\end{minipage}
\end{figure}

\begin{figure}
\begin{minipage}{0.5\textwidth}
    \includegraphics[width=8.5cm]{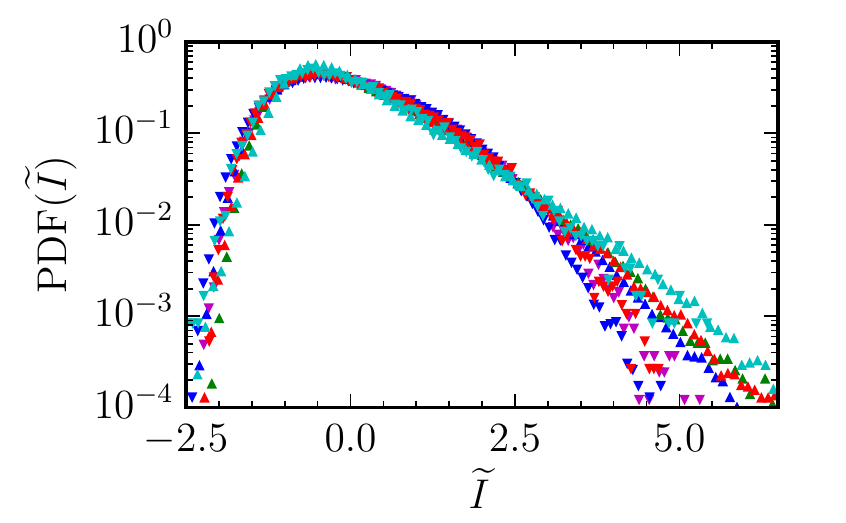}
    \caption{Rescaled histogram of all sampled ion saturation current time series, normalized according to
        \Eqnref{is_normalization}. Color coding of the
        plot markers is as in \Tabref{shotparams}, triangle up denotes data sampled at outboard
        mid-plane, triangle down denotes data sampled by the outermost divertor probe.}
    \label{fig:hist_is_collapsed}
\end{minipage}
\begin{minipage}{0.5\textwidth}
    \includegraphics[width=8.5cm]{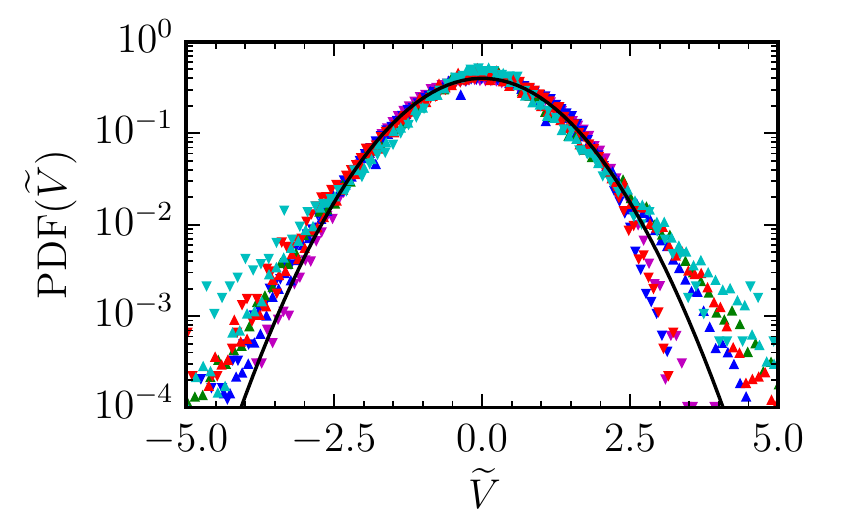}
    \caption{Rescaled histogram of all sampled floating potential time series, normalized according to
        \Eqnref{vf_normalization}. Color coding of the
        plot markers is as in \Tabref{shotparams}, triangle up denotes data sampled at outboard
        mid-plane, triangle down denotes data sampled by the outermost divertor probe.}
    \label{fig:hist_vf_collapsed}
\end{minipage}
\end{figure}

The histograms of the normalized burst amplitudes, \Figref{asp_amean} and 
\Figref{osd_amean}, are furthermore compatible with the assumption the pulse amplitudes 
are exponentially distributed. The evidence for this is however less clear than for 
the waiting times, mostly because the exponential model given by \Eqnref{PDF_expon} is 
only a good fit in the uppermost decade. This is due to the fixed location parameter and 
the conservation of probability limiting the choices of the slope of the cumulative 
distribution function. The estimated shape parameter of the stochastic model
\Eqnref{noisy_sn} is $1 \lesssim \gamma \lesssim 10$ for all distributions sampled in the
far scrape-off layer. This described the low intermittency case, i.e. pulses 
arrive frequently and overlap as to form large amplitude burst events. As a consequence 
the amplitudes taken from the bursts in the time series overestimate the underlying 
pulse amplitudes. This is reflected in the curved shape of the histograms 
\Figsref{asp_amean} and \ref{fig:osd_amean}. However, the presented maximum likelihood 
estimates agree well with the complementary cumulative distribution function over 
approximately one decade. 

\begin{table}
    \begin{tabular}{c|c|c|c}
        Shot    & $\neng$   &   \mean{A}                        & \mean{\tauw} \\ \hline
        $1$     & $0.15$    & $-$       / $0.37$    / $0.27$    & $-$       / $0.37$    / $0.64$    \\
        $2$     & $0.28$    & $0.12$    / $0.26$    / $0.38$    & $1.1$     / $0.61$    / $0.49$    \\
        $3$     & $0.32$    & $0.20$    / $-$       / $-$       & $1.2$     / $-$       / $-$       \\
        $4$     & $0.31$    & $0.26$    / $0.37$    / $0.43$    & $1.2$     / $0.66$    / $0.64$    \\
        $5$     & $0.42$    & $0.28$    / $0.37$    / $0.38$    & $1.3$     / $0.98$    / $0.91$    \\
    \end{tabular}
    \caption{Average amplitude and waiting times between conditionally averaged events compared
        for time series sampled by the horizontal scanning probe / divertor probe 9 / divertor probe 10.}
    \label{tab:ca_statistics_upstream_downstream}
\end{table}

Conditional averaging of the ion saturation current time series further reveals
an average burst shape that features a steep rise and a slow fall, both of which are 
well described by an exponential waveform. Typical rise times and fall times of the
events in the time series sampled by the horizontal scanning probe are given by
$\taur \approx 5\, \musec$ and $\tauf \approx 10\, \musec$, while the corresponding values
for the time series sampled by the divertor probes are larger by a factor of $2$.
We note however that the time resolution of the divertor probes is $2.5\, \mu \mathrm{s}$
which might affect the accuracy of the e-folding times negatively.
The conditionally averaged structure of the time series sampled by divertor
probe $10$ in discharge $5$ shows a larger asymmetry with a large fall time. However, 
the waveforms sampled by the divertor probes do not allow to draw conclusions about
the physical dimensions of impinging plasma filaments. The recorded waveform may
be due to both, a filament traversing the probe radially outwards or a filament impinging 
along the direction of the magnetic field on to the probe.

The conditionally averaged waveforms of the normalized ion saturation current and 
the floating potential signal, sampled at the outboard mid-plane far scrape-off layer, 
support the conventional picture of plasma blob propagation through the scrape-off layer. 
That is, peaks in the plasma particle density are associated with a dipolar electric 
potential structure whose polarization gives an electric drift velocity pointing towards 
the vessel wall. The phase shift between the conditionally averaged waveforms of the ion 
saturation current and floating potential is approximately $\pi / 2$ and the estimated 
radial velocities of the blobs structure are in the order of a few per cent of the ion 
acoustic velocity for all line averaged plasma densities. These results extend previous
measurements made in the scrape off layer of Alcator C-Mod \Ref{grulke-2006}.
A positive linear correlation is observed between the estimated radial blob velocities
and their normalized amplitude, with Pearson sample correlation coefficients given by
$0.19$ for $\neng = 0.28$ and $0.36$ for $\neng = 0.42$. A possible explanation for this 
correlation is that the pressure gradient within the blob structure increases with filament 
amplitude. Fluid modeling of isolated plasma filaments shows that the magnitude of the 
plasma pressure gradient increases the plasma vorticity associated with the plasma blob
\Ref{garcia-2006}. Assuming that the poloidal size of the plasma blobs is
constant \Ref{kube-2013}, this creates a larger electric field which in turn increases
the electric drift magnitude.

A simple estimate of parallel and perpendicular transport for a filament, see for example 
Fig. $3$ in \Ref{garcia-2007-coll}, suggests that the bursts in the time series sampled by 
the divertor probes may be due to plasma filaments impinging in the probes. For this we 
note that potential variations may also be caused by the internal temperature profile of 
plasma blobs \Ref{myra-2004}. For $\Te = 30\,\mathrm{eV}$ we evaluate the electron thermal
velocity to be $\vthe \approx 2 \times 10^{6}\, \mathrm{ms}^{-1}$ and the ion acoustic 
velocity in a deuterium plasma to be $\cs  = \sqrt{\Te / \mi} \approx 4 \times 10^{4}\, \mathrm{ms}^{-1}$. 
A lower bound on the characteristic velocity associated with transport of potential 
perturbations along the magnetic field is given by $\vthe$ \Ref{grulke-2014}.
Given a connection length of $L_\shortparallel \approx 10\, \mathrm{m}$, the particles
and potential generated by a blob ballooned on the low-field side will reach
the divertor targets after
$\tau_{\mathrm{n}, \parallel} \approx 3 \times 10^{-4}\, \mathrm{s}$
and
$\tau_{\mathrm{e}, \parallel} \approx 4 \times 10^{-6}\, \mathrm{s}$ respectively.
Assuming that the blob is instantiated as a structure with sharp modulation along a
flux tube and that it propagates normal to the flux tube at $\vrad = 500\, \mathrm{ms}^{-1}$,
independent of poloidal angle, the time estimate above implies that the footprint of
the blob has reached the divertor sheaths at radial coordinates given by
$\rho_{\mathrm{E}} \approx 2 \times 10^{-3}\, \mathrm{m}$ 
and
$\rho_{\mathrm{n}} \approx 1 \times 10^{-1} \mathrm{m}$.
This interpretation is compatible with previous results from correlation analysis of 
particle density proxies along single field lines in Alcator C-Mod and NSTX 
\Ref{grulke-2006, maqueda-2010, grulke-2014}.

Neglecting electron temperature fluctuations, a dipolar potential structure measured by 
the divertor probes may be interpreted as the footprint of a plasma blob. When this is 
the case, the radial velocity scaling of the plasma filaments falls in the sheath 
connected regime. On the other hand, a random waveform implies that the electric current 
loop within a plasma filament closes upstream of the divertor. On the other hand, random
structures have been observed in numerical simulations of plasma blobs where the late blob
has dispersed by Rayleigh-Taylor and Kelvin-Helmholtz instabilities \Ref{garcia-2006} and in
three-dimensional simulations, where no coherent structure of the late blob is recognizable 
\Ref{easy-2014}.

The hypothesis that the blobs are not electrically connected to the sheaths in discharge 5
is however also compatible with measurements of radial blob velocities in high density 
plasmas in Alcator C-Mod, which indicate that the radial filament velocity at the outboard 
mid-plane increases with increasing line-averaged density and exceed the value predicted 
for sheath connected blobs \Ref{agostini-2011, kube-2013}. 

\begin{figure}
\begin{minipage}{0.5\textwidth}
    \includegraphics[width=8.5cm]{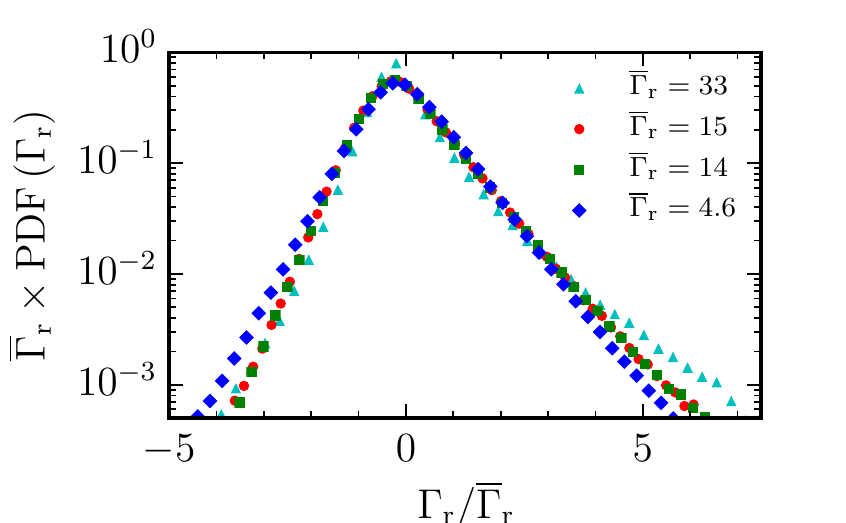}
    \caption{Rescaled histogram of the radial particle flux as sampled by the horizontal scanning probe.
        The particle flux is in units of $10^{21} \mathrm{m}^{-2} \mathrm{s}^{-1}$.}
    \label{fig:hist_flux_normalized}
\end{minipage}
\end{figure}

This hypothesis is further supported by histograms of the radial particle flux in the 
outboard mid-plane far scrape-off layer, shown in \Figref{hist_flux_normalized}. Upon 
proper normalization, the histograms for discharges $2$, $3$, and $4$ collapse, while 
the histogram for discharge $5$ features a more elevated tail. All histograms feature 
exponential tails for both the positive and the negative abscissa. The average radial 
particle flux increases by a factor of approximately $8$ from the discharge with 
$\neng = 0.28$ to the discharge with $\neng = 0.42$.  This increase of the radial
particle flux with line-averaged plasma density is consistent with previous experiments
in the Alcator C-Mod tokamak \Ref{labombard-2001, labombard-2002} as well as with
experiments performcd in the TCV tokamak \Ref{garcia-2007-jnm}. The higher frequency of
large flux events is consistent with the observation that blobs are moving faster while their 
cross-field size diameter remains constant \Ref{kube-2013}. Another possible explanation 
for the high average radial particle flux in discharge 5 may be increased levels of 
temperature fluctuations due to plasma filaments. 

The effective convective velocity, 
$\veff = \smean{\Gamma} / \smean{n_\mathrm{e}} = \smean{\widetilde{\Isat} \widetilde{\left( V^{\mathrm{SW}} - V^{\mathrm{NW}} \right)}} / B \triangle_\mathrm{p}$
, also increases with the line-averaged plasma density. 
For discharge 2 we find 
$\veff = 57 \mathrm{m} \mathrm{s}^{-1}$, while we find 
$\veff = 1.1 \times 10^{2} \mathrm{m} \mathrm{s}^{-1}$ for discharge 3 and 4. For discharge $5$
with $\neng = 0.42$ we find 
$\veff = 1.5 \times 10^{2} \mathrm{m} \mathrm{s}^{-1}$.

Such an increase in effective convective velocity is consistent with flux measurements
done in the Alcator C-Mod tokamak \Ref{labombard-2001}. The values we find for $\veff$ are
furthermore in the same order of magnitude as measured in the Tore Supra Device \Ref{fedorczak-2012}
and in TCV \Ref{garcia-2007-jnm}.

\section{Conclusion}
\label{sec:conclusion}
To conclude, we have studied the dependence of fluctuations in scrape-off layer plasmas
on the line-averaged particle density, as measured by Langmuir probes at the outboard 
mid-plane location and embedded in the outer divertor of Alcator C-Mod.
Time series of ion saturation current, sampled in the far-scrape off layer, all feature 
dynamics which is governed by the intermittent arrival of large amplitude burst events. 
Waiting times between large amplitude burst events are well described by an exponential 
distribution. Sub records of the time series feature a quadratic relation between 
coefficients of sample skewness and sample excess kurtosis. The fact that large-amplitude 
events occur uncorrelated and the quadratic relation between sample skewness and excess 
kurtosis, support assumptions of a stochastic model for the density fluctuations in 
scrape-off layer plasmas. The probability density function of this model describes 
the histograms of all sampled ion saturation current time series well.

The conditionally averaged waveform of the associated potential fluctuations is dipolar, 
except for time series sampled in the divertor plasma where the divertor is detached. This 
supports the hypothesis that plasma blobs are electrically detached in sufficiently high 
density plasmas and may explain the observed increase in radial blob velocity with 
line-averaged plasma density \Ref{agostini-2011, kube-2013}. Electric disconnection 
of the plasma blobs from the divertor sheaths may also explain recent experiments 
performed at the ASDEX Upgrade tokamak where it was observed that the radial blob 
velocity and cross-field diameter increases as the divertor detaches \Ref{carralero-2014}.

Future work will include a more detailed comparison of the stochastic model to time series 
measured in scrape-off layer plasmas. Work on a manuscript providing a detailed discussion 
of the stochastic model presented in this article is in progress.

\section{Acknowledgements}
R.K, O.E.G and A.T. were supported with financial subvention from the Research Council of Norway under 
grant 240510/F20. R.K. would like to thank D. Brunner for providing the script used to create
figure 1. Work partially supported by US DoE Cooperative agreement DE-FC02-99ER54512 at 
MIT using the Alcator C-Mod tokamak, a DoE Office of Science user facility. 
R.K and O.E.G. acknowledge the generous hospitality of the Plasma Science and Fusion Center at MIT
during a sabbatical stay during which these experiments were performed.

%%%%%%%%%%%%%%%%%%%%%%%%%%%%%%%%%%%%%%%%% Alcator C-Mod %%%%%%%%%%%%%%%%%%%%%%%%%%%%%%%%%%%%%%%%%%%%%%%%%%%%%%%%%%%%%%%%%%%%%%%
\newpage
\clearpage

\newpage
\clearpage

%%%%%%%%%%%%%%%%%%%%%%%%%%%%%%%%%%%%%%%% FSP: Te profiles %%%%%%%%%%%%%%%%%%%%%%%%%%%%%%%%%%%%%%%%%%%%%%%%%%%%%%%%%%%%%%%%%%%%%%%%
\newpage
\clearpage

%%%%%%%%%%%%%%%%%%%%%%%%%%%%%%%%%%%%%%%% Line-averaged density profiles and probe positions %%%%%%%%%%%%%%%%%%%%%%%%%%%%%%%%%%%%%%%%%%
\newpage
\clearpage
\centering

%%%%%%%%%%%%%%%%%%%%%%%%%%%%%%%%%%%%%%%%% Histogram fits %%%%%%%%%%%%%%%%%%%%%%%%%%%%%%%%%%%%%%%%%%%%%%%%%%%%%%%%%%%%%%%%%%%%%%%
% /Users/ralph/uni/cmod_paper/graphics/asp_histogram_fit.py
\newpage
\clearpage
\centering

% /Users/ralph/uni/cmod_paper/graphics/asp_histogram_fit.py
\newpage
\clearpage
\centering

%%%%%%%%%%%%%%%%%%%%%%%%%%%%%%%%%%%%%%%%%%%%%%
% /Users/ralph/uni/cmod_paper/gaphics/asp_histogram_fit.py
\newpage
\clearpage
\centering

% 
%%%%%%%%%%%%%%%%%%%%%%%%%%%%%%%%%%%%%%%% ASP average burst shape %%%%%%%%%%%%%%%%%%%%%%%%%%%%%%%%%%%%%%%%%%%%%%%%%%%%%%%%%%%%%%%
% /Users/ralph/uni/cmod_paper/graphics/ASP/asp_plot_burst_condavg_008.py
\newpage
\clearpage
\centering

%%%%%%%%%%%%%%%%%%%%%%%%%%%%%%%%%%%%%%%%% Scatter: radial velocity vs burst amplitude  %%%%%%%%%%%%%%%%%%%%%%%%%%%%%%%%%%%%%%%%%%%%%%%%%%%%
% /Users/ralph/uni/cmod_paper/graphics/asp_plot_burst_epol_scatter.py
\newpage
\clearpage
\centering

%%%%%%%%%%%%%%%%%%%%%%%%%%%%%%%%%%%%%%%% ASP waiting time distribution %%%%%%%%%%%%%%%%%%%%%%%%%%%%%%%%%%%%%%%%
\newpage
\clearpage
\centering

%%%%%%%%%%%%%%%%%%%%%%%%%%%%%%%%%%%%%%%% ASP burst amplitude histogram %%%%%%%%%%%%%%%%%%%%%%%%%%%%%%%%%%%%%%%%%
\newpage
\clearpage
\centering

%%%%%%%%%%%%%%%%%%%%%%%%%%%%%%%%%%%%%%%%% OSD Histogram fits %%%%%%%%%%%%%%%%%%%%%%%%%%%%%%%%%%%%%%%%%%%%%%%%%%%%%%%%%%%%%%%%%%%%%%%
%/Users/ralph/uni/cmod_paper/graphics/osd_histogram_fit_008.py
\newpage
\clearpage
\centering

%%%%%%%%%%%%%%%%%%%%%%%%%%%%%%%%%%%%%%%%% OSD Histogram fits %%%%%%%%%%%%%%%%%%%%%%%%%%%%%%%%%%%%%%%%%%%%%%%%%%%%%%%%%%%%%%%%%%%%%%%
%/Users/ralph/uni/cmod_paper/graphics/osd_histogram_fit_011.py
\newpage
\clearpage
\centering

%%%%%%%%%%%%%%%%%%%%%%%%%%%%%%%%%%%%%%%%% OSD average burst shape%%%%%%%%%%%%%%%%%%%%%%%%%%%%%%%%%%%%%%%%%%%%%%%%%%%%%%%%%%%%%%%%%%%%%%%
% /Users/ralph/uni/cmod_paper/graphics/osd_burst_shape.py
\clearpage
\centering

%%%%%%%%%%%%%%%%%%%%%%%%%%%%%%%%%%%%%%%%%% OSD Waiting time, burst amplitude distribution  %%%%%%%%%%%%%%%%%%%%%%%%%%%%%%%%%%%%%%%%
\newpage
\clearpage
\centering

\newpage
\clearpage
\centering

%%%%%%%%%%%%%%%%%%%%%%%%%%%%%%%%%%%%%%%%%% S vs K for ASP/OSD  %%%%%%%%%%%%%%%%%%%%%%%%%%%%%%%%%%%%%%%%%%%%%%%%%%%%%%%%%%%%
% /Users/ralph/uni/cmod_paper/graphics/asp_s_vs_k.py
\newpage
\clearpage
\centering

%%%%%%%%%%%%%%%%%%%%%%%%%%%%%% Normalized isat/Vfloat histograms %%%%%%%%%%%%%%%%%%%%%%%%%%%%%%%%%%%%%%%%%%%%%%%%%%%%%%%%%%
\newpage
\clearpage
\centering

%
%%%%%%%%%%%%%%%%%%%%%%%%%%%%%%%%%%%%%%%%%%%% Normalized radial flux PDF %%%%%%%%%%%%%%%%%%%%%%%%%%%%%%%%%%%%%%%%%%%%
\newpage
\clearpage
\centering

%%%%%%%%%%%%%%%%%%%%%%%%%%%%%%%%%%%%%%%% Bibliography %%%%%%%%%%%%%%%%%%%%%%%%%%%%%%%%%%%%%%%%%%%%%%%%
\newpage
%\bibliographystyle{apsrev4-1}
%\bibliography{cmod_refs}

\end{document}